\documentclass[twocolumn]{aastex631}

\usepackage{amsmath}
\usepackage{physics}
\usepackage{bbold}
\usepackage{soul}

\renewcommand{\thefootnote}{\fnsymbol{footnote}}

\newcommand{\review}[1]{#1}
\newcommand{\boldreview}[1]{#1}

\shortauthors{Habegger, Zweibel, \& Wong}

\received{2022-Nov-07}
\revised{2023-Apr-03}
\accepted{2023-Apr-21}
\submitjournal{The Astrophysical Journal}

\begin{document}

\title{The Impact of Cosmic Ray Injection on Magnetic Flux Tubes in a Galactic Disk}
\shorttitle{The Impact of Cosmic Ray Injection on Magnetic Flux Tubes in a Galactic Disk}


\correspondingauthor{Roark Habegger}
\email{rhabegger@wisc.edu}

\author[0000-0003-4776-940X]{Roark Habegger}
\affiliation{University of Wisconsin-Madison Astronomy Department }
\affiliation{University of Wisconsin-Madison Physics Department }

\author[0000-0003-4821-713X]{Ellen G. Zweibel}
\affiliation{University of Wisconsin-Madison Astronomy Department }
\affiliation{University of Wisconsin-Madison Physics Department }

\author[0000-0003-3133-6810]{Sherry Wong}
\affiliation{University of Wisconsin-Madison Astronomy Department }
\affiliation{University of Wisconsin-Madison Computer Sciences Department }

\begin{abstract}
\review{ In a seminal paper, Parker (1966) showed the vertical stratification of the interstellar medium (ISM) is unstable if magnetic fields and cosmic rays provide too large a fraction of pressure support. Cosmic ray acceleration is linked to star formation, so Parker's Instability and its nonlinear outcomes are a type of star formation feedback. Numerical simulations have shown the instability can significantly restructure the ISM, thinning the thermal gas layer and thickening the magnetic field and cosmic ray layer. However, the timescale on which this occurs is rather long $(\sim 0.4\,\mathrm{Gyr})$. Furthermore,  the conditions for instability depend on the model adopted for cosmic ray transport. In this work, we connect the instability and feedback problems by examining the effect of \review{a single, spatially and temporally localized} cosmic ray injection on the ISM over $ (\sim 1\,\mathrm{kpc}^3)$  scales. We perform cosmic ray magnetohydrodynamic simulations using the \texttt{Athena++} code, varying the background properties, dominant cosmic ray transport mechanism, and injection characteristics between our simulation runs. We find robust effects of buoyancy for all transport models, with  disruption of the ISM on timescales as short as $100$ Myr when the background equilibrium is dominated by cosmic ray pressure. }
\end{abstract}


\section{Motivation} 
\label{sec:Intro}
\review{
In 1966, Eugene Parker showed the interstellar medium's (ISM's) stratification was unstable when non-thermal pressure support by magnetic field and cosmic rays was introduced
\citep{1966Parker}. The free energy source for the instability is the gravitational potential energy of gas held above its natural scale height. Parker argued that undular perturbations
of the field lines, which on average are tangent to the Galactic plane, would allow gas to slide into the magnetic troughs, while the crests were free to buoyantly rise. The end result would be the production of cosmic ray inflated magnetic lobes rising perpendicular to the galactic disk and 
pockets of dense gas in the Galactic plane. 

In order for the instability to occur,  the  equation of state (EoS) of the medium must be soft enough to allow compression into pockets. Parker assumed the cosmic ray pressure would remain constant along field lines and that the gas would follow an isothermal EoS. Under these conditions, \textit{any} pressure support from magnetic fields and/or cosmic rays renders the system unstable.  }

The Milky Way's thermal energy density, magnetic energy density, and cosmic ray energy density are all on the order of $\sim 10^{-12} \,\mathrm{erg} \, \mathrm{cm}^{-3}$; e.g. \citep{2001Ferriere,2011Draine}. \review{In general, the far infrared - radio correlation \citep{2003Bell} suggests that star formation, cosmic ray acceleration, and magnetic field generation are linked. This correlation means the Parker instability connects star formation to the disruption of the ISM and the production of outflows. So, the Parker instability is a form of star formation feedback and could play a key role in galactic evolution. It could mediate the formation of dense clouds, but also heat and disrupt them \citep{2021Bustard}. It could launch mass from the disk as a wind or fountain \boldreview{\citep{2008Everett,2012Uhlig,2013Hanasz,2016Girichidis,2017Ruszkowski,2017WienerGlobal,2018Farber,2018Zhang,2018Hopkins,2019Chan,2021Hopkins}.} Perhaps less directly related to feedback, it could contribute to cosmic ray escape from the disk, and could influence the galactic dynamo by growing the component of the field perpendicular to the disk \boldreview{\citep{2000Hanasz,2004Hanasz,2009Hanasz}}}.

\review{
While the non-thermal pressure in cosmic rays and aforementioned buoyancy mechanism provide a way for cosmic rays to affect galactic evolution and the structure of the ISM, there is evidence against the Parker instability being a dominant effect in galaxies. The original Parker instability model has been revised to include more physical processes and more realistic setups \citep{1974Mouschovias,1975Zweibel,1978Asseo,1993Giz,2018Heintz}, and analyzed in the \boldreview{nonlinear} regime with numerical simulations \citep{2020Heintz,2022Tharakkal}. Overall, these results lead to more stringent conditions for instability and show the  instability requires a longer time to vertically restructure the  ISM; typically over $400\,\mathrm{Myr}$ \citep{2020Heintz,2022Tharakkal}. This long timescale means the effects of the instability could be overshadowed by other processes with shorter timescales.}

\review{One key difference between the different analyses lies in the treatment of cosmic ray transport and the choice of cosmic ray EoS. A stiffer adiabatic index ($\gamma_g$ (gas) or $\gamma_c$ (cosmic rays)) makes the system more stable; a softer EoS  is destabilizing. Parker's original assumption, $\gamma_c = 0$, makes the system very unstable \citep{2018Heintz}.} \review{Of course, not all cosmic ray effects can be expressed through an EoS, diffusion being an obvious exception. We defer a detailed discussion of transport to Section \ref{sec:Back:CRTransport}, but the main choices are between the self confinement model (also known as the streaming model), confinement by extrinsic turbulence,
which is diffusive, and the limit of zero diffusion (also known as advection). }

As a result of the EoS dependence on transport, the chosen transport model determines whether the cosmic rays stabilize or destabilize the system \citep{2017Zweibel,2018Heintz} In particular, cosmic rays are \textit{stabilizing} under the advection model. This result is puzzling, given the simple physical basis for cosmic ray buoyancy, and partly motivates  the work described in this paper. 

The fundamental question \review{connecting research on the Parker instability to research on star formation is, ``Under what conditions do cosmic rays change the ISM of a galaxy?'' To probe this question, we examine how a single localized injection of cosmic rays (e.g., particles accelerated in the shock waves created by  supernova explosions) disrupts the ISM. We take this route instead of a traditional Parker instability seeded by small velocity perturbations because it directly connects our setup to star formation feedback. } Localized cosmic ray injection has been studied previously as a driver for the galactic dynamo \boldreview{\citep{2000Hanasz,2004Hanasz,2009Hanasz}}. It produces structure similar to the Parker instability, with an extended magnetic lobe and gas condensed in parts of the disk. Focusing on the sources of cosmic rays instead of the background led us to ask the following: \review{``\textit{What is the timescale on which a single cosmic ray injection changes the vertical structure of the surrounding ISM?}"} We break this question into four sub-questions concerning \review{a single injection's impact on the ISM:}
\begin{itemize}
    \item (Q1) How does the background medium's stability change the injection's impact? 
    \item (Q2) How does the choice of cosmic ray transport model change the injection's impact?
    \item (Q3) How do the injection's strength and vertical location change its impact? 
    \item (Q4) How does cosmic ray injection differ from heat injection by a thermal explosion? 
\end{itemize} 
By answering these questions \review{for a range of models within an extended parameter space}, our simulations provide a detailed look at how \review{the injection of} cosmic rays \review{in a single event} changes the ISM on 
$\sim 1 \,\mathrm{kpc}$ \review{length scales and over times shorter than the $\sim 400\,\mathrm{Myr}$ required for restructuring by the Parker instability} \citep{2020Heintz,2022Tharakkal}. 

Overall, we show that for a stiff cosmic ray EoS, and transport by advection or diffusion, cosmic ray injection can \review{loft magnetic fields and cosmic rays to heights exceeding a kiloparsec} on timescales $\approx 100 \, \mathrm{Myr}$. Cosmic ray transport dominated by streaming ( \review{i.e. a strong magnetic field}) results in buoyant rising, producing structure similar to the Parker instability on timescales $\gtrsim 170 \, \mathrm{Myr}$. \review{ While injecting cosmic rays in a burst replicates some features of the Parker Instability, the medium responds on a timescale that is several times faster, and the dependence on transport is entirely different. Taken together our simulations show that cosmic ray injection can cause significant dynamical change on a $\sim 100\,\mathrm{Myr}$ timescale, with the change being largest if diffusion is the dominant cosmic ray transport process, the  magnetic field is weak, and the background cosmic ray pressure is large.} 

The paper is split into six sections including this motivation section. In Section \ref{sec:Back}, we provide background on the Parker instability and cosmic ray transport. In Section \ref{sec:Setup} we cover our numerical methods, our initial conditions for magnetohydrodynamic simulations, and our parameter choices for each simulation run. \review{In Section \ref{sec:Results} we present the overall results of our simulations.} In Section \ref{sec:Disc} we compare the runs and discuss the implications of our simulations, while answering our four questions concerning various astrophysical parameters. In Section \ref{sec:Conc} we summarize the paper and provide key takeaways from the work. Readers primarily interested in our results should focus on \review{Figures \ref{fig:TubePC} \& \ref{fig:Fast}, Table \ref{tab:Sims}, and Sections \ref{sec:Disc} \& \ref{sec:Conc}.}

In the appendices, we provide further detail on our simulations. \review{In Appendix \ref{app:Compare}, we compare a Parker instability simulation run with our code to one obtained previously with a significantly different code and show they give consistent results.} In Appendix \ref{app:Bound}, we discuss the boundary conditions and dimensionality of our simulations. In Appendix \ref{app:Convergence}, we discuss the dependence of our simulation results on various numerical parameters and our parameter choices for the primary simulations. 

\section{Background}
\label{sec:Back}

\subsection{Magnetohydrostatic Equilibrium}\label{sec:Back:HSE}
The Parker instability disrupts a magnetohydrostatic equilibrium in the galactic disk \cite{1966Parker}. \review{With a few exceptions
\citep{1978Asseo,1990Boulares}, }the equilibrium quantities are assumed to be functions of $z$ alone and the magnetic field lines are horizontal and straight. The condition for equilibrium is then
\begin{equation}\label{eq:hse}
 \frac{d}{dz}\left(P_g+P_c+\frac{B^2}{8\pi}\right)=\rho g.   
\end{equation}
\review{Parker solved the stability problem for a particularly simple class of equilibria: fixed, constant gravity $g$, constant mean squared random gas velocity $P_g/\rho$, and uniform ratios of nonthermal to thermal gas pressures parameterized by two constants $\alpha$ and $\beta$\footnote[3]{Note Parker's $\beta $ is \textbf{not} the \review{usual} plasma beta. Instead, the parameter $\alpha$ is the inverse of the plasma beta $\beta_\mathrm{pl} = P_g P_B^{-1}= \alpha^{-1}$.}}
\begin{equation}
    P_B = \frac{B^2}{8\pi} = \alpha P_g 
    \hspace{0.4in}
    P_c = \beta P_g .
    \label{eqn:Ratios}
\end{equation}
\review{While other features of the equilibrium setup were modified in later work, the $(\alpha,\beta)$ parameterization has generally endured, and we use it to describe the pre-injection state in our simulations. However, we use the equilibrium setup from \cite{1993Giz} as our initial ISM background. In this model the gravitational profile is smooth} 
\begin{equation}
    \vb{g}(z) = - g_* \hat{z}\tanh\left( \frac{z}{H_*} \right).
    \label{eqn:TanhGravity}
\end{equation}
instead of the discontinuous $-g_*\mathrm{Sign}(z)$ profile \cite{1966Parker} used. The smooth function in Equation \ref{eqn:TanhGravity} poses fewer numerical difficulties than a discontinuous profile \review{in numerical simulations which extend above and below the galactic plane}  \citep{2020Heintz}. In Equation \ref{eqn:TanhGravity}, the asymptotic vertical gravitational acceleration $g_*>0$ and the gravitational scale height $H_*$ depend on the structure of the galactic disk's stellar population.  

Assuming a plane-parallel hydrostatic equilibrium, \review{Equation \ref{eq:hse} becomes}
\begin{equation}
    \frac{d P_\mathrm{tot}}{d z}\equiv
    (1+\alpha+\beta) \frac{d P_g}{d z} = \rho(z) g(z)  
    \label{eqn:HSE}
\end{equation} 
where $P_\mathrm{tot}$ is the total pressure; the sum of gas, magnetic, and cosmic ray pressure. We solve for the equilibrium under the influence of the gravitational profile Equation \ref{eqn:TanhGravity}, using an isothermal equation of state $P_g = c_s^2 \rho$ with constant sound speed $c_s$, and midplane values $P_g(0) = P_{g,0},\, \rho(0) = \rho_0$. The solution to Equation \ref{eqn:HSE} is 
\begin{equation}
    \frac{P_g(z)}{P_{g,0}} = \frac{\rho(z)}{\rho_{0}}  = f(z) = \mathrm{sech}^{\eta}{\left[ \frac{z}{\eta H} \right]}.
    \label{eqn:HSEsol}
\end{equation}
The solution depends on how the scale height of the gas $H$ relates to the gravitational scale height. The scale height of the gas is
\begin{equation}
    H = \frac{c_s^2}{g_*} (1+\alpha + \beta).
    \label{eqn:ScaleH}
\end{equation}
The ratio of the two scale heights $\eta = H_* / H$ is a constant. Taking a limit of Equation \ref{eqn:HSEsol} as $\eta \rightarrow 0$ (or equivalently, $H_* \rightarrow 0 $) recovers the solutions in \cite{1966Parker}. 

\begin{figure}[b]
    \includegraphics[width=0.48\textwidth]{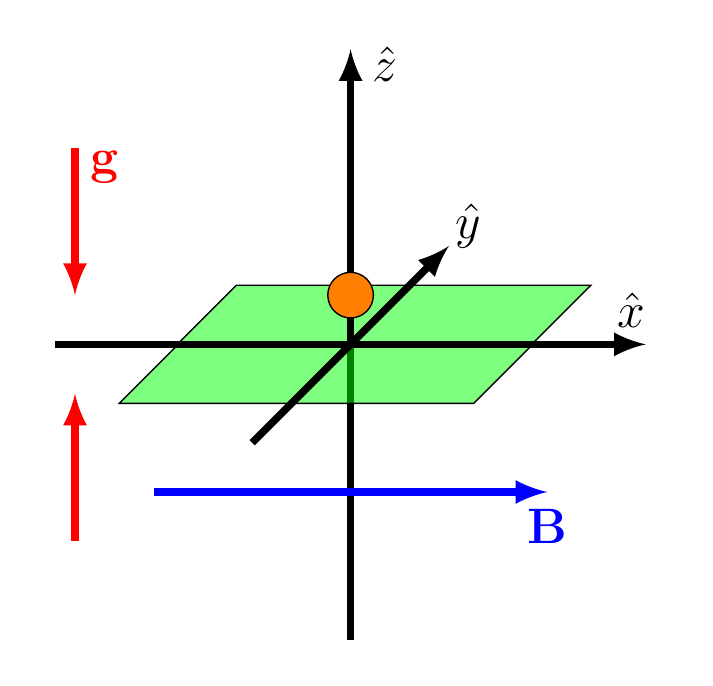}
     \caption{This schematic shows our initial setup for hydrostatic equilibrium. We have a gravitational field (red arrows) in the $\hat{z}$ direction, which flips sign at the midplane $z=0$ (see Equation \ref{eqn:TanhGravity}). The magnetic field (blue arrow) is oriented along the $\hat{x}$ direction and creates a magnetic pressure supporting vertical hydrostatic equilibrium (Equations \ref{eqn:HSE},\ref{eqn:MagField}). We then inject cosmic rays, shown by an \review{orange} circle (see Section \ref{sec:Setup:Initial} for injection profile), above the midplane, which is shown as a green rectangle.}
    \label{fig:Setup}
\end{figure}

\review{The geometry and coordinate system for this} equilibrium are illustrated in Figure \ref{fig:Setup}. The pressures, including magnetic pressure, are homogeneous in surfaces parallel to the $xy$ plane, which is shown as a green plane in Figure \ref{fig:Setup}. For our simulations, we orient the initial magnetic field in the $\hat{x}$ direction. The magnetic field is shown as a blue arrow in Figure \ref{fig:Setup}. Using Equations \ref{eqn:Ratios} \& \ref{eqn:HSEsol} the magnetic field is 
\begin{equation}
    \vb{B}(z) = \hat{x} \sqrt{8\pi  \alpha P_{g,0} f(z)}.
    \label{eqn:MagField}
\end{equation}
Similarly, the cosmic ray pressure is 
\begin{equation}
    P_c(z) = \beta P_{g,0} f(z).
    \label{eqn:CRP}
\end{equation}

In Figure \ref{fig:Setup}, the coordinates $(\hat{x},\hat{y},\hat{z})$ approximately map to a galactic disk's cylindrical coordinates. The $\hat{x}$ direction is parallel or anti-parallel to the azimuthal direction $\hat{\phi}$ (depending on whether the magnetic field is oriented clockwise or counter-clockwise around the galactic center), the $\hat{y}$ direction is parallel or anti-parallel to the radial direction $\hat{r}$ (again, depending on magnetic field orientation), and the $\hat{z}$ direction is parallel to the cylindrical $\hat{z}$ direction. 

\subsection{Parker Instability}
\label{sec:Back:Parker}

We can assess the stability of \review{Parker's equilibrium to ideal (energy conserving) small amplitude perturbations with the generalized Schwarzschild} convection criterion \citep{1961Newcomb,1990Boulares,2017Zweibel}. This criterion for instability in a vertically stratified atmosphere is 
\begin{equation}
    -\frac{d \ln \rho}{dz} < \frac{\rho g_*}{\gamma_g P_g}.
    \label{eqn:Newcomb}
\end{equation}
\review{This criterion is applicable above the midplane with $z>0$. Assuming we are well above the midplane $(z > H_*)$ where the gravitational acceleration is constant $(-g_*)$, then the logarithmic derivative is the scale height $H$ of the gas. This assumption gives the same criterion as Parker's original gravity profile. In this limit, the criterion for instability to occur is }
\begin{equation}
     \alpha + \beta > \gamma_g - 1 .
     \label{eqn:OldParker}
\end{equation}
If we include the compressibility of cosmic rays, then we need to replace $\gamma_g P_g$ with $\gamma_g P_g+ \gamma_c P_c$ in Equation \ref{eqn:Newcomb}. Using this change and evaluating the criterion again, we get 
\begin{equation}
     \alpha + \beta (1 - \gamma_c) > \gamma_g - 1 .
     \label{eqn:NewParker}
\end{equation}
The above criterion illustrates a confusing result noted in \cite{2017Zweibel}, and again in \cite{2018Heintz}. Generically, one would guess increasing non-thermal pressures makes the system more unstable. However, Equation \ref{eqn:NewParker} shows there is a way for increasing cosmic ray pressure $(\beta)$ to make the system more stable. Since the cosmic ray fluid is relativistic gas, it has $\gamma_c = \frac{4}{3}$.  In that case, the left hand side would be smaller with increasing $\beta$. \review{The physical explanation for this effect is that compressing the cosmic rays requires work. }

\boldreview{Self confinement by streaming leads to the polytropic relation $P_c\propto\rho^{2/3}$ along magnetic flux tubes. Using this value in Equation \ref{eqn:NewParker} would predict a stability threshold intermediate between ``classic" Parker ($\gamma_c=0$) and $\gamma_c=4/3$. Instead, it is shown through a modal analysis in \cite{2018Heintz} that self confinement leads to a even larger domain of instability and faster growth rates than ``classic" Parker with $\gamma_c = 0$. In fact, the analysis used to derive Equation \ref{eqn:Newcomb} does not apply to the streaming model, due to the relative drift between cosmic rays and thermal gas, and the heating that accompanies streaming. Similarly, Equation \ref{eqn:Newcomb} does not apply to diffusion dominated cosmic ray transport. }

\begin{figure}
     \includegraphics[width=0.48\textwidth]{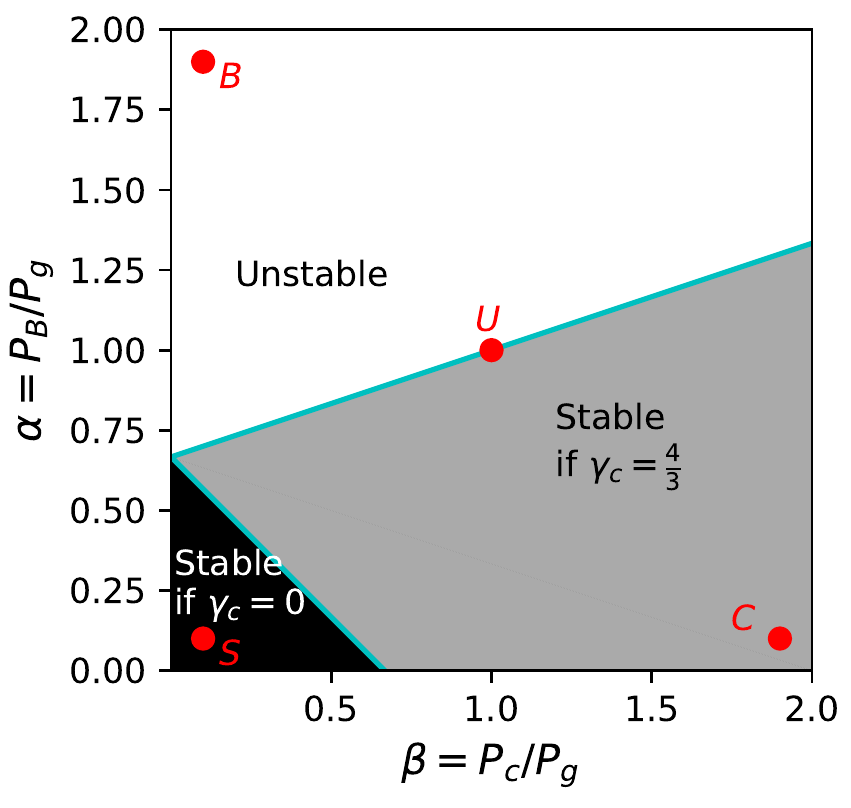}
        \caption{This schematic shows the parameter space of magnetic pressure $\alpha = P_B / P_g$ and cosmic ray pressure $\beta = P_c / P_g$. For advective transport, we show the regions of parameter space where the initial hydrostatic equilibrium (see Section \ref{sec:Back:Parker}) is unstable to the Parker instability. The gray region is stable if cosmic rays have an adiabatic index $\gamma_c = 4/3$. The black region is stable if the cosmic rays are an infinitely compressible fluid with index $\gamma_c = 0$. The S, U, B, and C points are where the simulations, listed in Table \ref{tab:Sims}, are located in this parameter space.}
        \label{fig:ParkerStability}
\end{figure}

\review{We show a schematic for advection dominated transport} in Figure \ref{fig:ParkerStability}. This figure is not quantitatively predictive of our simulation setup, \review{which has a smooth $g_*$ given by Equation \ref{eqn:TanhGravity}, a non-ideal perturbation, and in some cases, transport dominated by streaming or diffusion.} The schematic shows the $(\alpha,\beta)$ space of equilibrium solutions with an adiabatic gas exponent $\gamma_g = 5/3$. Using Equations \ref{eqn:OldParker} and \ref{eqn:NewParker}, we show different regions of stability determined by the cosmic ray adiabatic index $\gamma_{c}$. There is a strictly Parker unstable region (unshaded), a strictly Parker stable region (black), and \review{a gray region which is stable if the cosmic rays take work to compress (assuming $\gamma_c=4/3$).} For smaller values of $\gamma_g$, the fan of stability boundaries shifts downward so the boundaries intersect on the vertical axis where $\alpha = \gamma_g-1$. 

\boldreview{To reiterate: We choose simulation parameters throughout the $(\alpha,\beta)$ parameter space to probe different background medium conditions. This short review of stability of the background medium illustrates how our different background medium models would be classified in a Parker instability analysis, and provides a short review of the Parker instability and its dependence on cosmic ray transport.}

\subsection{Cosmic Ray Transport}
\label{sec:Back:CRTransport}
The stability of the system described in Section \ref{sec:Back:Parker} and illustrated in Figure \ref{fig:ParkerStability} depends on the effective compressibility of the cosmic rays, and therefore on cosmic ray transport. Cosmic rays are transported throughout the interstellar medium by three mechanisms: advection, \review{diffusion, and streaming. \boldreview{The effects of these different transport mechanisms have already been examined in global galactic models \citep{2016Girichidis,2017WienerGlobal}. } 

All three mechanisms are based on the assumption that the cosmic rays exchange momentum and energy by scattering from fluctuations in the magnetic field with a wavelength of order the cosmic ray gyroradius (so-called gyroresonant scattering). Advection, which applies in the limit of infinitely short mean free path and scattering by fluctuations with no preferred propagation direction along the background magnetic field lines,} is the most straightforward: if the thermal gas has a bulk flow in a particular direction, the cosmic rays should flow with the gas. If advection is dominant, then the cosmic rays \review{behave like a relativistic fluid, with negligible inertia, following the flow of the thermal (non-relativistic) gas.} 

Diffusion of cosmic rays \review{applies when the fluctuations again have no preferred propagation direction, but the mean free path to scattering is large enough to} allow the cosmic rays to leak through the thermal gas. \review{If the measured residence time and vertical scale height of} cosmic rays in the Milky Way \review{are explained by diffusive propagation, the implied diffusion coefficient is} $\kappa_c \sim 3 \cdot 10^{28} \,\mathrm{cm}^2 \,\mathrm{s}^{-1}$. 

Streaming\review{, or self confinement, applies when the magnetic fluctuations are generated by the cosmic rays themselves as a result of directional anisotropy in the frame of the fluctuations. It can be shown that in the short mean free path regime, the streaming direction is down the cosmic ray pressure gradient.} This instability appears in \review{the} kinetic theory for cosmic ray transport, and it drives the cosmic rays into a bulk flow at the Alfven speed $\vb{v}_A\equiv B/\sqrt{4\pi\rho}$ \citep{1969Kulsrud,1969Wentzel,1975Skilling}. The streaming instability also heats the gas: the cosmic rays transfer energy to hydromagnetic waves through gyro-resonance, which then dissipate energy into the thermal gas \citep{1969Kulsrud,1969Wentzel,1975Skilling}. This heating appears as a source term in the equation for the thermal energy density and the cosmic ray energy density. The \review{heating rate from the streaming instability is $ \vb{v}_A \cdot \grad P_c $ (see \cite{2017Zweibel} for discussion and references).} 

Each of these transport methods drives a different characteristic response in the ISM when cosmic rays are injected. Advection and streaming both drive steep fronts of gas and cosmic ray energy. \review{Whereas advection will result in the cosmic ray fluid moving at the flow speed, bringing the gas with it, streaming allows the cosmic rays to move ahead of the gas if the Alfven speed is faster than the sound speed. In the streaming case, a front of cosmic rays will move through the gas while heating the gas according to the Alfvenic heating term $ \vb{v}_A \cdot \grad P_c $.} Diffusion, in contrast to both streaming and advection, drives a smoother flow of cosmic rays while also exerting a force on the gas along the cosmic ray pressure gradient. 

We provide the following useful timescales for each transport mechanism to move cosmic rays through a distance $L$ of the ISM:
\begin{equation}
    \tau_\mathrm{adv} = L \cdot \left( \frac{\gamma_g P_0}{\rho_0} \right)^{-1/2}\approx 65 L_\mathrm{kpc}T_4^{-1/2}\rm{Myr},
    \label{eqn:timeAdv}
\end{equation}
\begin{equation}
    \tau_\mathrm{str} =  L \cdot  \left( \frac{B_0}{\sqrt{4\pi \rho_0}} \right)^{-1}\approx 35 L_\mathrm{kpc}\rho_{-24}^{1/2}B_{\mu}\rm{Myr}, 
    \label{eqn:timeStr}
\end{equation}
\begin{equation}
    \tau_\mathrm{diff} = \frac{L^2}{\kappa_c} \approx 30 L_\mathrm{kpc}^2\kappa_{28}^{-1}\rm{Myr}, 
    \label{eqn:timeDiff}
\end{equation}
\review{where $L_\mathrm{kpc}$, $T_4$, $B_{\mu}$, $\rho_{-24}$, and $\kappa_{28}$ are the length scale, gas temperature, gas density (assuming full ionization), and cosmic ray diffusivity in units of $1\,\mathrm{kpc}$, $10^4\,\mathrm{K}$, $1\mu\mathrm{G}$, $10^{-24} \,\mathrm{g}\,\mathrm{cm}^{-3}$, and $10^{28}\,\mathrm{cm}^{2}\,\mathrm{g}^{-1}$. }
In our simulations, the gas flows are generally subsonic, so the advection timescale is an imperfect measure: the correct timescale will be larger. For advection dominated simulations, we set the simulation parameters such that the streaming and diffusion timescales are much larger than an order unity multiple of the advection timescale in Equation \ref{eqn:timeAdv}. However, when the advection and streaming timescales are similar according to Equations \ref{eqn:timeAdv}\,\&\,\ref{eqn:timeStr}, the streaming transport dominates the dynamics in our simulations because of this overestimated flow speed. We only use these timescales to determine the dominant transport mechanism, defaulting to streaming when the streaming and advection timescales are similar (i.e. Simulation \texttt{U} - see Table \ref{tab:Sims}).

\section{Setup}
\label{sec:Setup}

\subsection{\review{Basic Equations and} Numerical Methods}
\label{sec:Setup:NumMethods}
\review{Our simulations are based on fluid equations for the thermal gas - cosmic ray - magnetic field system \cite{1991Breitschwerdt,2017Zweibel}. These equations accommodate cosmic ray transport by both diffusion and streaming; for the latter, it is assumed that the scattering waves propagate down the cosmic ray pressure gradient. Enforcing this feature creates numerical difficulties near extrema in cosmic ray pressure. Although this can be handled through a smoothing length \citep{2010Sharma}, we adopt the two moment formulation of streaming transport developed by \cite{2018Jiang} (see also  \cite{2021Thomas}) as implemented in the Athena++ code \citep{2020Stone}, which modifies the equations presented in \cite{1991Breitschwerdt} through a time dependent evolution equation for the cosmic ray flux.}

\review{The implementation of \texttt{Athena++} from \cite{2018Jiang} solves the following equations:}
\begin{equation}
    \pdv{\rho}{t}+ \div{\left( \rho \vb{v}\right)} = 0,
    \label{eqn:Continuity}
\end{equation}
\begin{equation}
    \pdv{\vb{B}}{t}
    - \curl{\left( \vb{v} \cross \vb{B} \right) }
    = 0,
    \label{eqn:Induction}
\end{equation}
\begin{multline}
    \pdv{ \big( \rho \vb{v}\big) }{t}
    + \div {\left( \rho \vb{v}\vb{v}
    + \left( P_g + \frac{1}{2} B^2 \right)\vb*{\mathbb{1}}
    - \vb{B}\vb{B} \right)} \\
    =  \rho \vb{g} + \vu*{\sigma}_{c} \vdot
    \left( \vb{F}_{c}
    - \gamma_c \vb{v}  E_c 
    \right),
    \label{eqn:JiangMomentum}
\end{multline}
\begin{multline}
    \pdv{E}{t} 
    + \div { \left( \vb{v} \left( E + P_g + \frac{1}{2}B^2 \right) - \vb{B} \left(\vb{B}\vdot \vb{v} \right) \right) } \\
    = \rho \vb{g} \vdot \vb{v} - (\gamma_c - 1) \left( \vb{v} + \vb{v}_s\right) \vdot \grad{ E_{c}} ,
    \label{eqn:JiangEnergy}
\end{multline}
\begin{equation}
    \pdv{E_c}{t} 
    + \div{ \vb{F}_{c}}
    = (\gamma_c - 1)\left( \vb{v} + \vb{v}_s\right) \vdot
    \grad{E_c} ,
    \label{eqn:JiangCREnergy}
\end{equation}
\begin{equation}
    \frac{1}{V_m^2} \pdv{\vb{F}_c}{t}
    + (\gamma_c - 1) \grad{ E_{c}}
    = -\vu*{\sigma}_{c} \vdot
    \left( \vb{F}_{c}
    - \gamma_c \vb{v} E_c 
    \right).
    \label{eqn:JiangCRMomentum}
\end{equation}
The dynamical variables are gas density $\rho$, bulk flow velocity $\vb{v}$, gas pressure $P_g$, magnetic field $\vb{B}$, the combined internal and kinetic energy density $E = P/(\gamma_g-1) + \rho v^2/2$, the cosmic ray energy density $E_c = P_c / (\gamma_c - 1)$, and the cosmic ray energy flux $\vb{F}_{c}$. \review{We include the effects of a gravitational acceleration $\vb{g}$, given by Equation \ref{eqn:TanhGravity}. We use $\gamma_c = 4/3$ and set the modified speed of light parameter to $V_m = 0.3c$. For more discussion on convergence according to this parameter, see Appendx \ref{app:Convergence}.}

\subsection{Initial Conditions}
\label{sec:Setup:Initial}
The initial profiles of density, gas pressure, magnetic field, and cosmic ray pressure are described in Section \ref{sec:Back}. We use midplane density and pressure values of $\rho_0 = 10^{-24} \, \mathrm{g} \,\mathrm{cm}^{-3}$ and $P_{g,0} = 10^{-12} \, \mathrm{g} \,\mathrm{cm}^{-3}$. We use an asymptotic gravitational acceleration $g_* = 4 \cdot 10^{-9} \, \mathrm{cm} \, \mathrm{s}^{-2}$ and a gravitational scale height $H_* = 100 \, \mathrm{pc}$ \citep{2020Heintz}. \review{This height is smaller to what is used in Parker instability simulations (see App.\,\ref{app:Compare} and \cite{2016Rodrigues,2020Heintz}). Additionally, this height decreases our initial profile's height ratio $\eta$ when compared to Milky Way and Parker instability simulations \citep{2016Rodrigues,2020Heintz}. This adjustment means our injection can occur in a region where the gravity has nearly reached the asymptotic value $g_*$. As the flux tube of interest rises, it will move in a constant gravitational acceleration.} 

The simulation grid is $200 \times 50 \times 300 \,\mathrm{cells}$ (ordered with respect to $(\hat{x},\hat{y},\hat{z})$) . The cell size is $10 \times 20 \times 10 \,\mathrm{pc}^3$, giving a total simulation volume $2 \times 1 \times 3 \,\mathrm{kpc}^3$. The \review{third} dimension $(\hat{z})$ is off-center of the midplane ($z=0$), extending from $z = -1 \,\mathrm{kpc}$ to $z = +2 \,\mathrm{kpc}$. The cosmic rays are injected above the midplane ($z_0>0$), so this off-centering focuses the computational resources on the injection and the resulting flows. We do not use any adaptive or static mesh refinement. \review{We adopt outflow boundary conditions in the $\hat{x}$ and $\hat{y}$ directions and vacuum (diode) boundary conditions in the $\hat{z}$ direction to minimize the effect of domain size. See Appendix \ref{app:Bound} for additional discussion.

The equilibrium is setup to within single-precision floating point numerical error, which is less than the density and pressure floors of our simulations. There are some waves created by the interaction of the equilibrium with the vertical boundaries where there is an abrupt transition to vacuum. These waves are small and have no long term effect on the dynamics of our simulation.}

The cosmic ray energy density injection profile is a 3D Gaussian function:
\begin{equation}
    \varepsilon_{c}(x,y,z) =
    \frac{E_\mathrm{SN}}{\left( 2\pi r_\mathrm{SN}^2 \right)^{3/2}} \cdot \exp \left[ \frac{-1}{2r_\mathrm{SN}^2} \left| \vb{r} - \vb{r}_0\right|^2\right].
    \label{eqn:InjectionProfile}
\end{equation}
The parameters of the injection are its position $\vb{r}_0$, radius $r_\mathrm{SN}$, and total energy injected $E_\mathrm{SN}$. We add this perturbation onto the background cosmic ray pressure profile in Equation \ref{eqn:HSEsol}. Integrating the profile, Equation \ref{eqn:InjectionProfile}, over volume gives $E_\mathrm{SN}$ as the total energy injected. The injection occurs at different heights above the midplane depending on the other parameters under consideration (see Table \ref{tab:Sims}). 

In each simulation, we use the same injection radius $r_\mathrm{SN} = 50 \, \mathrm{pc}$. This radius is large for a single supernova shock, \boldreview{but reasonable for our default energy injection from $\sim 10$ supernovae} (see Section \ref{sec:Setup:Justify}). This larger radius allows our simulation (with $10-20 \,\mathrm{pc}$ resolution) to reliably sample the injection profile. To avoid sampling errors (aliasing) which will change the total injected energy, we center the injection in the $xy$ plane on a cell center. The grid has cell faces along $x=0$ and $y=0$ planes, so we choose to center the injection at $x_0=5 \,\mathrm{pc}$ and $y_0=10 \, \mathrm{pc}$. As a result of this placement, the peak of the profile occurs in the center of a computational cell. The height $z_0$ and total energy $E_\mathrm{SN}$ are varied across our simulations (see \ref{tab:Sims})

\boldreview{For reference, if the pressure of $10^{51}\,\mathrm{erg}$ injected cosmic rays were uniformly distributed throughout the volume of the injection tube (radius $50\,\mathrm{pc}$, length $2\, \mathrm{kpc}$) is $\sim 0.73\times 10^{-12}$ dynes cm$^{-2}$, corresponding roughly to a doubling of background cosmic ray pressure. In this strictly 1D situation, with no horizontal pressure gradient, the tube would expand radially. This expansion would reduce its density and cause it to float upward buoyantly, with a characteristic rise time $\sim 100 \, \mathrm{Myr}$ (for our adopted gravitational field and scale height).} \newline

\begin{table*}
    \begin{tabular}{llll|l}
         Parameters & & & & Timescales \\\hline
         $\rho_0 = 10^{-24} \, \mathrm{erg} \, \mathrm{cm}^{-3}$ 
         & $H_* = 100 \, \mathrm{pc} $ 
         & 
         & $V_{m}= 0.3 c $ 
         & $\tau_\mathrm{adv} = 76 \, \mathrm{Myr} \left( \frac{\gamma_g}{5/3}\right)^{-1/2} $ \\
         $P_0 = 10^{-12} \, \mathrm{erg} \, \mathrm{cm}^{-3}$
         & $g_* = 4 \cdot 10^{-9} \, \mathrm{cm} \, \mathrm{s}^{-2}$ 
         & 
         & $\kappa_{c,\perp} = 3 \cdot 10^{24} \frac{\mathrm{cm}^2}{\mathrm{s}}$
         & $\tau_\mathrm{str} = 69 \, \mathrm{Myr} \left( \alpha\right)^{-1/2}  $ \\
         $c_s =  10^6 \, \mathrm{cm} \, \mathrm{s}^{-1} $
         & $r_\mathrm{SN} = 50 \, \mathrm{pc}$ 
         & 
         & $\gamma_c = 4/3$ 
         & $\tau_\mathrm{diff} = 10 \, \mathrm{Myr} \left( \frac{\kappa_c}{10^{28} \frac{\mathrm{cm}^2}{\mathrm{s}}}\right)^{-1} $\\
         $T = 1.21 \cdot 10^4 \, \mathrm{K}  \, (\bar{m} /m_p) \, $
         & $B_0 = 5.06 \, \mu\mathrm{G} \sqrt{\alpha}$ 
         & 
         & 
         & \\
    \end{tabular}
    \begin{center}
    \begin{tabular}{|l||r|r|r|r|r||r|r|r||r|r||r|} \hline
        Run Name 
        & $\alpha$ 
        & $\beta$
        & $\gamma_g$
        & $H\left( \mathrm{pc} \right)$
        & $\eta$
        & $z_0\left( \mathrm{pc} \right)$
        & Weight $\left( M_\odot\right)$
        & $E_\mathrm{SN} \left( \mathrm{erg} \right)$
        & CR
        & $\kappa_{c,\parallel}\left( \mathrm{cm}^2\, \mathrm{s}^{-1}\right)$
        & \review{Question} \\\hline
        \texttt{S} 
        & $0.1$ & $0.1$ & $5/3$ 
        & $97$
        & $1.03$
        & $45$
        & $1.28 \cdot 10^{4}$
        & $10^{51}$
        & Adv.
        & $3 \cdot 10^{24}$
        & Q1 \\\hline
        \texttt{U} 
        & $1.0$ & $1.0$ & $5/3$ 
        & $243$
        & $0.41$
        & $105$
        & $2.24 \cdot 10^{4}$
        & $10^{51}$
        & Str.
        & $3 \cdot 10^{24}$
        & Q1,Q3,Q4 \\\hline
        \texttt{Uiso} 
        & $1.0$ & $1.0$ & $1.1$ 
        & $243$
        & $0.41$
        & $105$
        & $2.24 \cdot 10^{4}$
        & $10^{51}$
        & Str.
        & $3 \cdot 10^{24}$
        & Q1 \\\hline \hline
        \texttt{Cadv} 
        & $0.1$ & $1.9$ & $5/3$ 
        & $243$
        & $0.41$
        & $105$
        & $2.24 \cdot 10^{4}$
        & $10^{51}$
        & Adv.
        & $3 \cdot 10^{24}$
        & Q1,Q2 \\\hline
        \texttt{Bstr} 
        & $1.9$ & $0.1$ & $5/3$ 
        & $243$
        & $0.41$
        & $105$
        & $2.24 \cdot 10^{4}$
        & $10^{51}$
        & Str.
        & $3 \cdot 10^{24}$
        & Q1,Q2 \\\hline
        \texttt{Cdiff} 
        & $0.1$ & $1.9$ & $5/3$ 
        & $243$
        & $0.41$
        & $105$
        & $2.24 \cdot 10^{4}$
        & $10^{51}$
        & Diff.
        & $3 \cdot 10^{28}$
        & Q1,Q2 \\\hline 
        \texttt{Bdiff} 
        & $1.9$ & $0.1$ & $5/3$ 
        & $243$
        & $0.41$
        & $105$
        & $2.24 \cdot 10^{4}$
        & $10^{51}$
        & Diff.
        & $3 \cdot 10^{28}$
        & Q2 \\\hline \hline
        \texttt{Uheavy} 
        & $1.0$ & $1.0$ & $5/3$ 
        & $243$
        & $0.41$
        & $25$
        & $3.27 \cdot 10^{4}$
        & $10^{51}$
        & Str.
        & $3 \cdot 10^{24}$
        & Q3 \\\hline
        \texttt{Ublast} 
        & $1.0$ & $1.0$ & $5/3$ 
        & $243$
        & $0.41$
        & $105$
        & $2.24 \cdot 10^{4}$
        & $10^{52}$
        & Str.
        & $3 \cdot 10^{24}$
        & Q3 \\\hline
        \texttt{Utherm} 
        & $1.0$ & $1.0$ & $5/3$ 
        & $243$
        & $0.41$
        & $105$
        & $2.24 \cdot 10^{4}$
        & $10^{51\,*}$
        & Str.
        & $3 \cdot 10^{24}$
        & Q4 \\\hline 
    \end{tabular}
\end{center}
    \caption{The first table shows the
    parameters held fixed across our simulations, as well as derived timescales. The second table shows the parameters we vary for each simulation run we perform. The first column shows each simulation run name. The next columns show background medium parameters: magnetic pressure to gas pressure ratio $\alpha$, cosmic ray pressure to gas pressure ratio $\beta$, thermal adiabatic index $\gamma_g$, gas scale height $H$, and profile exponent $\eta$ (see Equation \ref{eqn:HSEsol}). The second set of columns shows injection parameters: initial height of the injection $z_0$, the weight in the column above the injection, and the volume integrated injection energy $E_\mathrm{SN}$ (see Equation \ref{eqn:InjectionProfile}). The third set of columns shows dominant cosmic ray transport mechanism and the parallel cosmic ray diffusion coefficient $\kappa_{c,\parallel}$. The final column shows which questions the simulation applies to (see Sections \ref{sec:Intro}\,\&\,\ref{sec:Setup:Justify}).}
    \label{tab:Sims}
\end{table*}

\subsection{\review{Explanation} of Parameters}
\label{sec:Setup:Justify}
\review{ We present the results of ten simulations of cosmic ray injection. The initial parameters for each simulation are shown in Table \ref{tab:Sims}.} We choose the parameters for the simulations to help us answer the four questions, presented at the end of Section \ref{sec:Intro} and labeled Q1, Q2, Q3, and Q4. 

We first considered what values of $\alpha$ and $\beta$ are reasonable in the \review{solar neighborhood}. Measurements suggest the pressures (thermal, cosmic ray, and magnetic) are nearly equal \citep{2001Ferriere}. Therefore, we choose a base run with $\alpha=\beta=1$. Simulations with these values are named with a prefix \texttt{U}. In Figure \ref{fig:ParkerStability}, this is on a stability boundary when cosmic ray \review{streaming and diffusion are} not included. However, each simulation in Table \ref{tab:Sims} includes transport by diffusion (even when it is not dominant), and most include streaming. Therefore, the stability implied by Figure \ref{fig:ParkerStability} does not strictly apply, suggesting this equal pressure background medium is Parker unstable. To compare this background with a strictly Parker stable background, we run a simulation with $\alpha=\beta=0.1$. We use a prefix `S' to refer to this combination of values. Our first two simulations, labelled \texttt{U} and \texttt{S}, allow us to \review{address Q1: }how the stability of the background medium changes the effect of cosmic ray injection. For both these simulations, we use a gas adiabatic constant consistent with a monatomic ideal gas $\gamma_g = 5/3$, a negligible cosmic ray diffusion constant $\kappa_c = 3\cdot 10^{24} \,\mathrm{cm}^2 \, \mathrm{s}^{-1}$, and an injection energy equivalent to the \review{estimated cosmic ray injection energy of $\sim 10$ supernovae \citep{2014Caprioli}, such as might be expected from an association of coeval massive stars.}

Our third simulation\review{, Uiso}, has the same parameters as \texttt{U} except for the adiabatic gas constant $\gamma_g=1.1$, which brings the system closer to the isothermal case originally considered by \cite{1966Parker} \review{and is expected to be more unstable. This simulation also addresses Q1.}

To answer Q2, \review{how cosmic ray transport affects the behavior of the system post injection,} we need a way to differentiate the dominant cosmic ray transport mechanism. We \review{do this} by considering each transport timescale, given by Equations \ref{eqn:timeAdv},\,\ref{eqn:timeStr},\,\ref{eqn:timeDiff}. \review{It turns out that the} flows in our simulations are subsonic and our advection timescale assumes propagation at the sound speed. So, when streaming and advection timescales are close, we assume streaming dominates. Simulations \texttt{U} and \texttt{S}, are instances of streaming and advection dominance, respectively. 

To \review{better} isolate the impact of each transport mechanism, we probe two other points in the $(\alpha,\beta)$ plane of Figure \ref{fig:ParkerStability}. We run a simulation with a large magnetic field $\alpha=1.9$ and low amount of background cosmic rays $\beta=0.1$, so it is in the top left corner of Figure \ref{fig:ParkerStability}. This simulation is labelled \texttt{Bstr}, and streaming dominates the transport of the injected cosmic rays because the Alfven speed is much higher than the flow speed and the diffusion rate. We also run a simulation \texttt{Cadv} with $\alpha = 0.1$ and $\beta =1.9$, placing it in the bottom right hand corner of Figure \ref{fig:ParkerStability}.  With this simulation, we probe a \review{medium which is} Parker stable \review{due to the stiff equation of state} with large non-thermal pressure where streaming \review{and diffusion are} subdominant transport mechanisms \review{ relative to advection.}. The simulation \texttt{Cdiff} has the same values of $\alpha$ and $\beta$, but uses a diffusion constant $\kappa = 3 \cdot 10^{28} \, \mathrm{cm}^2 \, \mathrm{s}^{-1}$  which is close to estimated Milky Way values. 

While these three simulations allow us to \review{examine} how streaming, advection, and diffusion \boldreview{affect the response of the medium to cosmic ray injection, they do not completely eliminate the influence of the background state; while the nonthermal pressures are equal in all three, the role of magnetic tension in \texttt{Bstr} is larger than in \texttt{Cadv} or \texttt{Cdiff} due to the stronger background magneic field. Therefore, to determine whether the streaming cosmic rays or the strong magnetic field was more critical, we ran \texttt{Bdiff}. That simulation had the same initial parameters as \texttt{Bstr}, but with a higher diffusion coefficient and streaming terms removed. Although diffusion and advection do not have an `off-switch' in the implementation from \cite{2018Jiang}, streaming transport does, and we make use of it when running \texttt{Bdiff}. }

\texttt{Bstr} and \texttt{Cadv} \review{also address} Q1 because they are at different points in the $(\alpha,\beta)$ plane. They have different background cosmic ray and magnetic pressures from simulations \texttt{U} or \texttt{S}, allowing us to learn how that aspect of the background medium changes the impact of cosmic ray injection. 

\review{Question 3 probes the effects of injection properties, so} we adjust parameters related to the injection. Using simulation \texttt{U} as a control case, we compare \review{both of the following simulations} to it, and only change single parameters. First, we use simulation \texttt{Uheavy}  with the cosmic ray injection at a lower height in the disk, meaning it has more weight above it. Second, we use a simulation, \texttt{Ublast}, with the strength of $10\%$ of $100$ supernovae to understand how different amounts of energy injection change the results. Other than the weight above the injection and the injection energy, we keep the variables the same as simulation \texttt{U}, which we compare each of these cases against. 

For Q4, \review{which compares the effects of cosmic rays to direct energization of thermal gas,} we run a simulation, \texttt{Utherm}, with the same initial parameters as the Parker unstable simulation \texttt{U}, except the injection is in thermal pressure instead of in cosmic ray pressure. 

\noindent
\begin{figure*}
    \centering
    \includegraphics[width=0.98\textwidth]{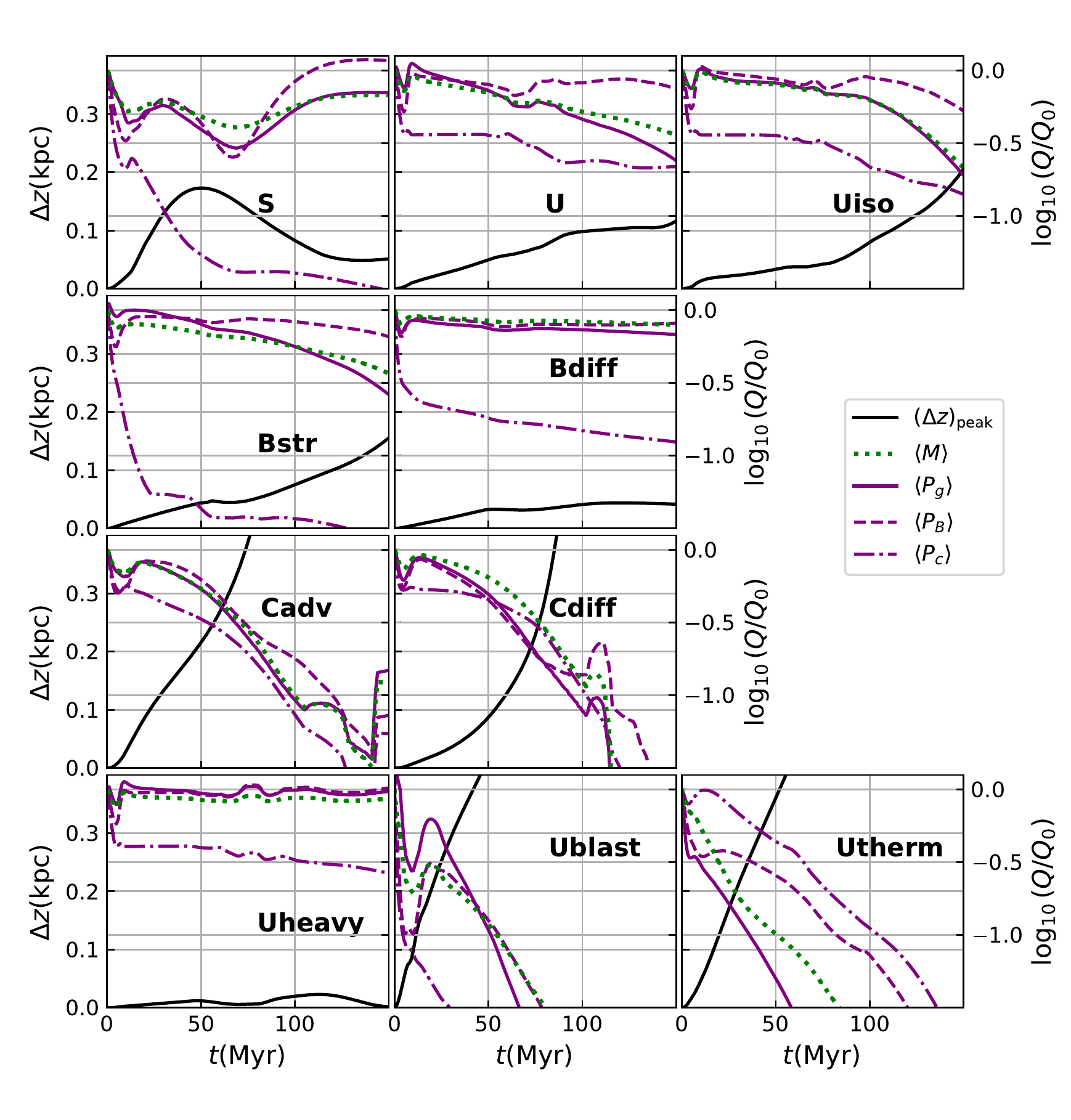}
    \caption{\review{After injecting cosmic rays on a magnetic flux tube, we track that tube's movement. For each simulation shown, the solid black line is maximum height of the flux tube at a given time. The green dotted line is the average mass in the region $x<|250\,\mathrm{pc}|$, as a fraction of the initial average mass. The purple solid, dashed, and dash-dotted lines are the gas, magnetic, and cosmic ray pressure, respectively, averaged in the same region. The second $y$ axis on the right side applies to the lines of mass and pressures, with each quantity $Q \in \{ \langle M \rangle, \langle P_g \rangle , \langle P_B \rangle, \langle P_c\rangle\}$ normalized by either $P_0=10^{-12}\,\mathrm{erg}\,\mathrm{cm}^{-3}$ or the initial value (for mass $\langle M \rangle$). }
    }
    \label{fig:AllSims}
\end{figure*}
\section{Results}
\label{sec:Results}

We focus on a flux tube enclosing the initial injection. Since we use the ideal MHD equations and because there is effectively no perpendicular cosmic ray diffusion, the gas and cosmic rays stay on the flux tube. \review{All the simulations share some common features. The cosmic rays in the injection sphere are overpressured by more than a factor of 20. This creates a low density cavity within a few Myr and launches a pressure driven flow away from the injection sphere. The cavity itself is buoyant because of reduced density and the outflow reduces the density along a progressively longer segment of the tube. The result is a rising magnetized arch which becomes ever more buoyant as gas is drained from the tube by a gravity driven downflow. The only forces that counter the rise are magnetic tension due to field line curvature, the inertia of the overlying gas, and adiabatic cooling of both the thermal and cosmic ray gases as they expand. The configuration has strong up-down asymmetry, because the effects which slow the flux tube's rise increase near the midplane.}

\review{For each simulation, we examine physical quantities near the center of the tube $(x=\pm250\,\mathrm{pc})$. This restriction allows us to focus on the dynamics in the center of the flux tube, where the initial cosmic ray injection occurred. Additionally, it minimizes the impact of the boundary conditions on our results for early times in the simulation (See Appendix B for a discussion of boundary conditions). In Figure \ref{fig:AllSims} we show the maximum height along the flux tube as a solid black line, the average gas pressure as a solid purple line, the average magnetic pressure as a dashed purple line, the average cosmic ray pressure as a dash-dotted purple line, and the average mass as a dotted green line. The average flux tube height in the $\pm250\,\mathrm{pc}$ region is approximately the same as the maximum height.} These quantities help us understand how the injection evolves through time and how it causes the flux tube to change. 

\review{The most striking part of the results in Figure \ref{fig:AllSims} are the plots for simulations \texttt{Cadv}, \texttt{Cdiff}, \texttt{Ublast}, and \texttt{Utherm}. In these four simulations, the flux tubes rise over $400\,\mathrm{pc}$ in less than $100\,\mathrm{Myr}$. The rise is accompanied by decreases in mass and pressure in the central region of the flux tube. 

The other simulations do not exhibit such a quick, drastic rise in flux tube height, nor decreases in pressure and mass in the central region of the flux tube. Simulations \texttt{S} and \texttt{Uheavy} end up being stable to the perturbation (meaning the flux tube rise is limited). \texttt{S} even finds a new equilibrium after an initial rise. Simulations \texttt{U}, \texttt{Uiso}, and \texttt{Bstr} all eventually begin to buoyantly rise, but only rising to $200\,\mathrm{pc}$, at a time of $150\,\mathrm{Myr}$ after the cosmic ray injection. These simulations each have streaming as the dominant cosmic ray transport - instead of creating a violent disruption in the center of the magnetic flux tube, streaming puts energy into heating the entire length of the magnetic flux tube. However, the stronger injection in \texttt{Ublast} still launches material rapidly when compared to \texttt{U}, even though streaming is dominant. }

\begin{figure*}
    \centering
    \includegraphics[width=0.9\textwidth]{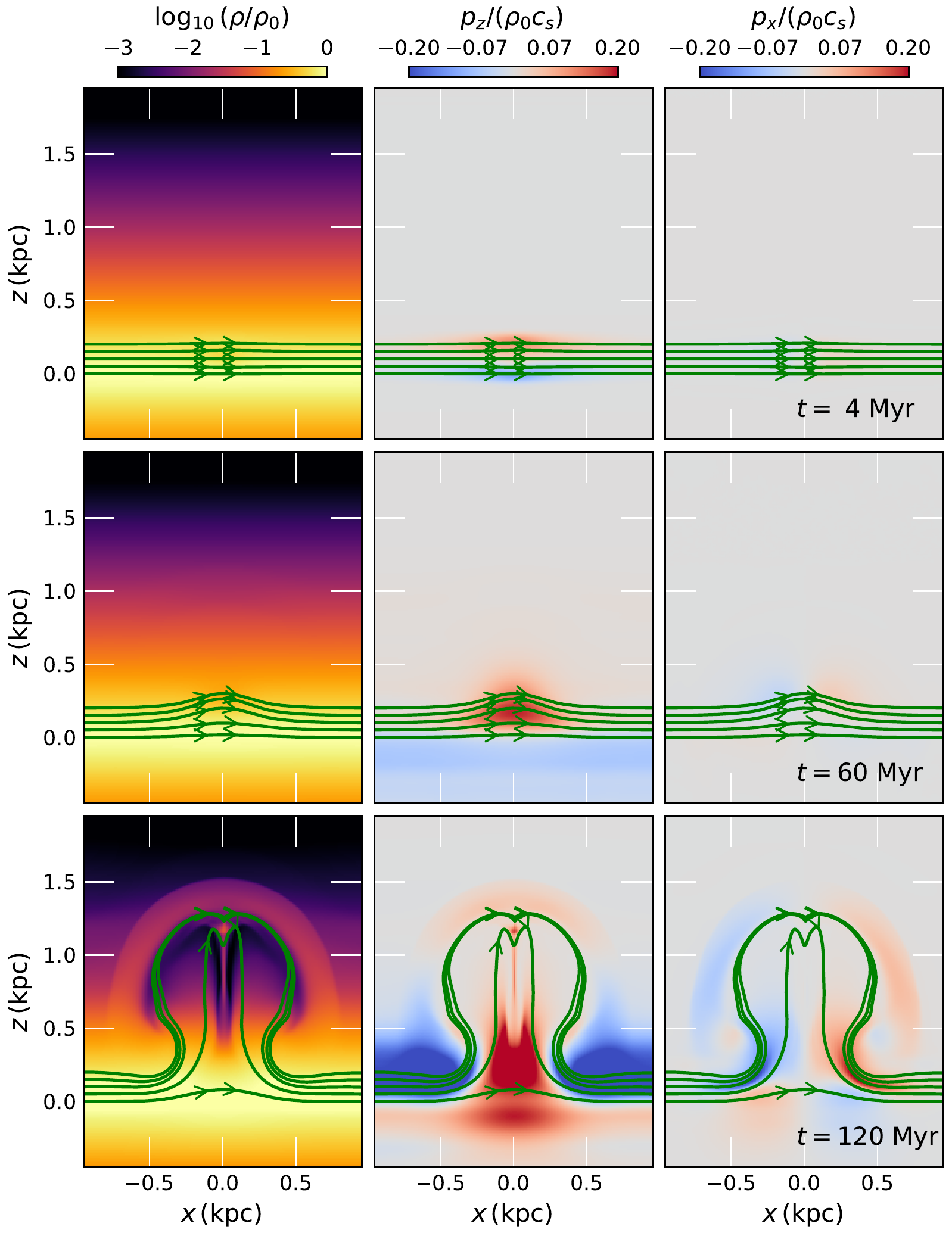}
    \caption{\review{Plots of gas density (first column), vertical momentum (second column), and horizontal ($\hat{x}$) momentum  (third column) at a $y=0$ cut of simulation \texttt{Cdiff}.} In this simulation, cosmic ray diffusion is the dominant transport mechanism. Early in the simulation, the cosmic rays quickly diffuse along the magnetic field lines. This diffusion leaves behind a flux tube with increased average cosmic ray pressure. This region begins to rise buoyantly, shown in the middle row. The final row shows gas falling back toward the disk around the injection, while the location continues to launch gas upward at $x=0$. The normalization constants are $P_0 = 10^{-12} \,\mathrm{erg} \,\mathrm{cm}^{-3}$ and $\rho_0 c_s = 10^{-18} \,\mathrm{g} \,\mathrm{cm}^{-2} \, \mathrm{s}^{-1}$. \review{The green lines are magnetic field lines; the middle line is integrated from the $x=-1\,\mathrm{kpc}$ bound at the height of the initial injection. The other lines are initialized at $\pm r_\mathrm{SN}$ and $\pm 2 r_\mathrm{SN}$ of the injection height.}}
    \label{fig:TubeCdiff}
\end{figure*}

\review{To more clearly illustrate the dynamics in our simulations, we show 2D cuts from simulation \texttt{Cdiff} in Figure \ref{fig:TubeCdiff}. The first column shows gas density, the second column shows vertical momentum, and the third column shows horizontal momentum. The green lines are magnetic field lines along the flux tube analyzed in Figure \ref{fig:AllSims}. The first row is a snapshot at $4\,\mathrm{Myr}$, the second row at $60\,\mathrm{Myr}$, and the third row at $120\,\mathrm{Myr}$. Once the cosmic rays diffuse, they create an over-pressured flux tube which begins to move upward (middle column, top row). After the flux tube bends, gas begins to fall down the curved lines (right column, middle row). After some time, this process builds until a large up-flow in the center and down-flow along the magnetic field lines. The final density plot (left column, bottom row) also shows how the flux tube is able to lift some mass out of the disk.}

\begin{figure*}
    \centering
    \includegraphics[width=0.9\textwidth]{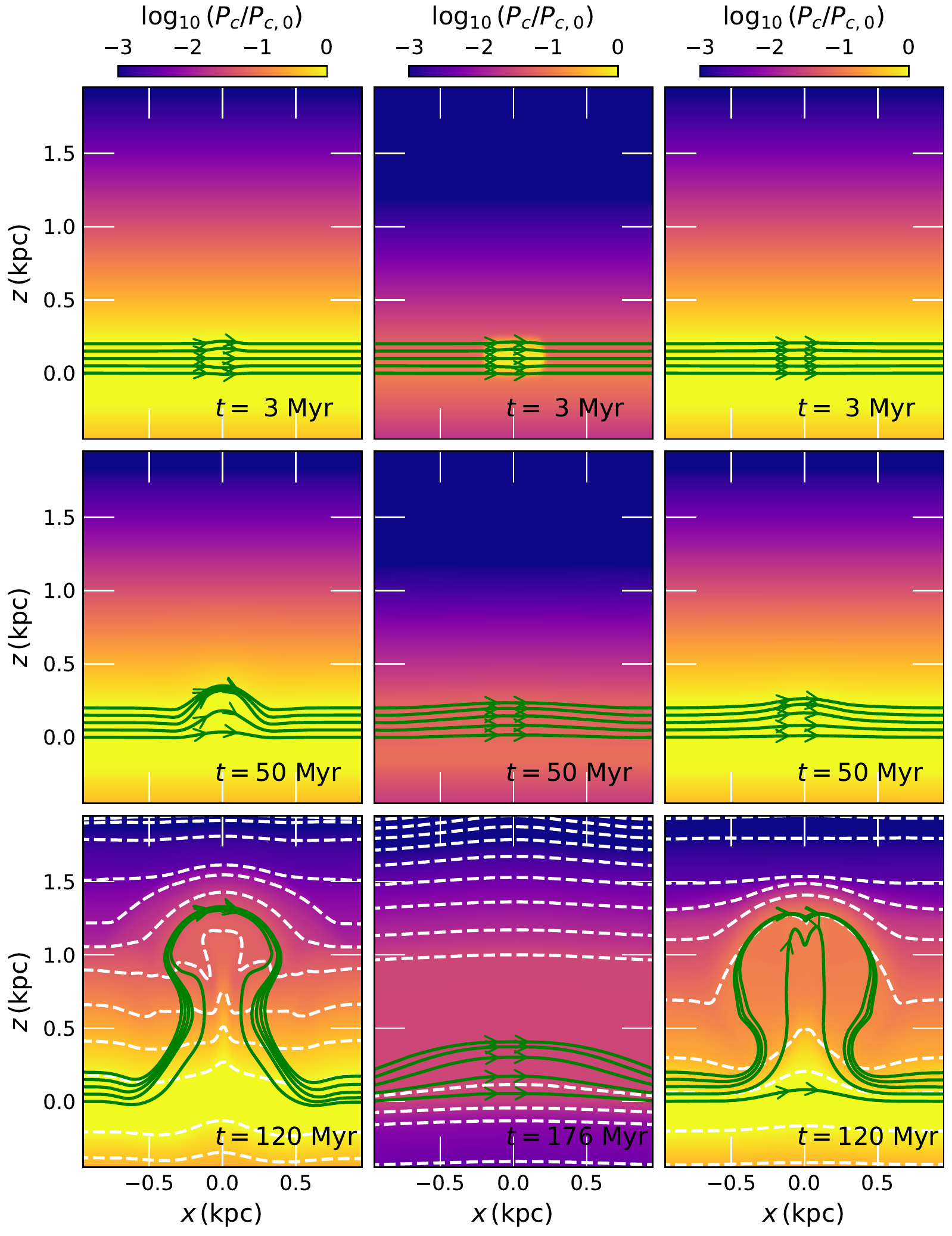}
    \caption{Plots of cosmic ray pressure at a $y=0$ cut of simulations \texttt{Cadv} (first column), \texttt{Bstr} (second column), and \texttt{Cdiff} (third column). The normalization constant is $P_0 = 10^{-12} \,\mathrm{erg} \,\mathrm{cm}^{-3}$. The green lines are magnetic field lines. The dashed white lines are contours of equal cosmic ray pressure. In the first two rows, cuts are all from the same time snapshots, but the final row shows simulation \texttt{Bstr} at a later time snapshot than the other two simulations.}
    \label{fig:TubePC}
\end{figure*}

\review{
We show the cosmic ray pressure distributions for simulations \texttt{Cadv}, \texttt{Bstr}, and \texttt{Cdiff} in Figure \ref{fig:TubePC}. The first column shows simulation \texttt{Cadv}, the second column shows \texttt{Bstr}, and the third column shows \texttt{Cdiff}. The first and second rows show the same time dumps for each simulation (top row is $3\,\mathrm{Myr}$, middle row is $50 \,\mathrm{Myr}$). The final row shows the later development of the injection for each simulation. The \texttt{Bstr} simulation does not grow as much vertically because magnetic tension holds the flux tube down for most of the simulation. Only once the field lines are able to bend does the flux tube begin to rise, because gas is able to leave the flux tube at a faster rate. This effect is also shown in the second row of plots in Figure \ref{fig:AllSims}. Regardless of cosmic ray transport by streaming or diffusion, the strong magnetic field simulations (\texttt{Bstr} and \texttt{Bdiff}) exhibit slowed or negligible rise of the flux tube. Streaming appears to be the most effective in the large magnetic field case because the gas gets heated in addition to being overpressured. }
\noindent

\noindent


\subsection{\review{A Caveat} on Simulation \emph{\texttt{Utherm}}}
\label{sec:Results:Utherm}
\boldreview{
In simulation \texttt{Utherm}, instead of injecting cosmic rays, we injected energy as thermal heating. This thermal energy injection had the same magnitude, $10^{51}\,\mathrm{erg}$, as the cosmic ray injection in simulation \texttt{U}.  }

\boldreview{
While this simulation highlights the effectiveness of thermal energy injection, our simulations lack radiative cooling. This injected gas should cool by emitting radiation while it expands. We estimate a radiative cooling time scale to determine the effect of radiative cooling on the injection, using Equation 34.4 from \cite{2011Draine}:
\begin{equation}
    t_\mathrm{cool} =1.1 \cdot 10^{5} \,\mathrm{yr}
    \left(\frac{T}{10^6 \,\mathrm{K}} \right)^{1.7}
    \left(\frac{n_\mathrm{H}}{\mathrm{cm}^3}\right)^{-1}
\end{equation}
For simulation \texttt{Utherm}, the peak temperature in the center of the injection is $10^{5.04}\,\mathrm{K}$, compared to the background $T_0 \sim 10^{4}\,\mathrm{K}$ Therefore, the radiative cooling time for this injection is $t_\mathrm{cool} \approx 2.2 \,\mathrm{kyr}$. This time scale is so short that this thermal injection should dissipate before causing a disruption in the ISM like we see in simulation \texttt{Utherm}. While temperatures are even higher in actual supernova remnants ($\sim 10^{7}\,\mathrm{K}$ for the Sedov-Taylor phase \citep{2011Draine}), our injection represents a long term average impact of those remnants. Radiative cooling will limit the impact of the thermal injection, as compared to a cosmic ray injection. A similar effect is seen in models of supernova heated gas launched from galactic disks, which tend to be fountains rather than winds \citep{1976Shapiro,1995Rosen,2018Bustard,2020Bustard}. 
}

\section{Discussion}
\label{sec:Disc}

To the questions posed in Section \ref{sec:Intro}, we now have the following answers: 
\begin{itemize}
    \item (Q1) A purely Parker stable medium limits the effect of cosmic ray injection. An isothermal medium is more prone to disruption by \review{a cosmic ray injection}. See Figure \ref{fig:Q1}.
    \item (Q2) \boldreview{Cosmic ray advection and diffusion drive changes in the ISM on \review{$\sim 100$ Myr} time scales, through explosive launching and buoyancy, respectively. \review{ Large magnetic field strength (which also implies streaming dominated transport) delays the flux tube's rise, but eventually causes buoyant rising. In the large magnetic field case, streaming is more disruptive than diffusion.} See Figure \ref{fig:Q2}.}
    \item (Q3) Stronger injections drive more explosive flows, and injections closer to the midplane take longer to launch the flux tube. See Figure \ref{fig:Q34}.
    \item (Q4) Thermal injection drives buoyant rising of the flux tube on a shorter time scale than cosmic ray injection. Cosmic rays decrease the average mass along the flux tube at a slower rate. \review{Eventually, cosmic ray injections overtake thermal injections in height. However, simulation \texttt{Utherm} overestimates the efficacy of thermal injection because it lacks cooling (see Section \ref{sec:Results:Utherm}).} See Figures \ref{fig:Q34},\,\ref{fig:Fast}.
\end{itemize}

In the following subsections, we provide more complete explanations for these answers. 

\subsection{Dependence on Background Medium (Q1)}
\label{sec:Results:Q1}

\begin{figure}
    \centering
    \includegraphics[width=0.48\textwidth]{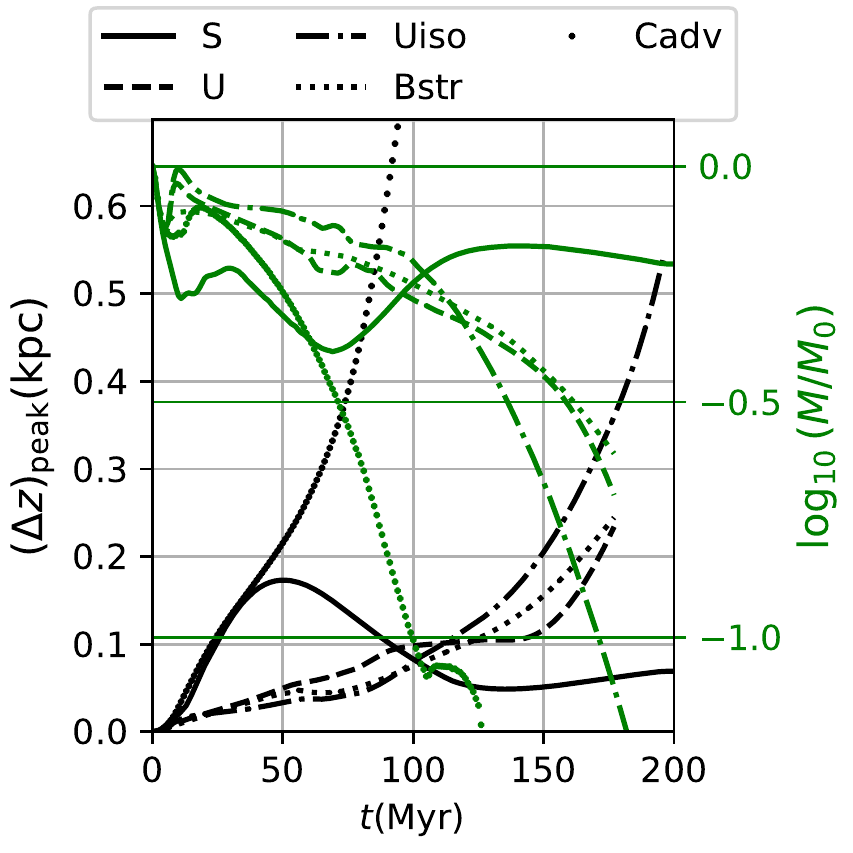}
    \caption{Plot of maximum tube height (black) and average mass (green) \review{ within $x = \pm 250 \,\mathrm{pc}$ of the injection location} for simulations related to Q1. Note the vertical axis has been extended compared to Figure \ref{fig:AllSims}. \review{Solid lines show simulation \texttt{S}, dashed lines show simulation \texttt{U}, dash-dotted lines show simulation \texttt{Uiso}, dotted lines show simulation \texttt{Bstr} and dots show simulation \texttt{Cadv}.} The streaming dominated simulations, \texttt{U} and \texttt{Bstr}, rise late. The advection dominated simulations, \texttt{Cadv} and \texttt{S}, have different results: \review{the medium with more cosmic ray pressure is significantly disrupted.} Simulation \texttt{Uiso} with $\gamma_g = 1.1$ allows more mass loss later in the simulation, when compared with simulation \texttt{U}, which had $\gamma_g = 5/3$.}
    \label{fig:Q1}
\end{figure}

We explored how changes in the background medium affected the evolution of a cosmic ray injection. The results are shown in Figure \ref{fig:Q1}, which plots the average mass and height along the flux tube of each simulation.

The Parker stable simulation (Simulation \texttt{S}) weathers the cosmic ray injection, at least under transport by advection and streaming, reaching a new stable equilibrium. When the background is Parker unstable according to the original criterion (Equation \ref{eqn:OldParker}), the injection drives significant change. For the simulations \texttt{U}, \texttt{Bstr}, and \texttt{Cadv}, we can use Equation \ref{eqn:OldParker} to show the combined nonthermal pressures dominate the gas compressibility $(\alpha + \beta = 2) > (\gamma_g - 1 = 2/3)$. The simulations with a background biased towards magnetic pressure (\texttt{Bstr}) or cosmic ray pressure (\texttt{Cadv}) launch the flux tube faster than the equipartition case (\texttt{U}). Therefore, the initial nonthermal pressures, given by $(\alpha,\beta)$, determine how prone the system is to disruption by cosmic ray injection. This dependence also suggests an ISM with non-uniform $\alpha, \beta$ would provide a complex environment for cosmic ray injection, since different directions and positions could be more (or less) prone to being disrupted. 

The case of an isothermal-like background with $\gamma_g  =1.1$ responds on a slightly shorter timescale than a background medium with $\gamma_g = 5/3$. This conclusion comes from considering the average height achieved by the flux tubes in simulations \texttt{Uiso} and \texttt{U}. In simulation \texttt{Uiso}, buoyant rising begins after $80\,\mathrm{Myr}$ and stays significant through the end of the simulation. For simulation \texttt{U}, buoyant rising becomes dominant only after $140\,\mathrm{Myr}$. The average mass also decreases at a faster rate at late times in simulation \texttt{Uiso}. Because $\gamma_g$ is lower, it takes less work to compress the gas and push it off the flux tube once buoyancy kicks in. This change allows mass to flow at a faster rate, driving the buoyancy force to become larger. 

\subsection{Dependence on Cosmic Ray Transport (Q2)}
\label{sec:Results:Q2}

\begin{figure}
    \centering
    \includegraphics[width=0.48\textwidth]{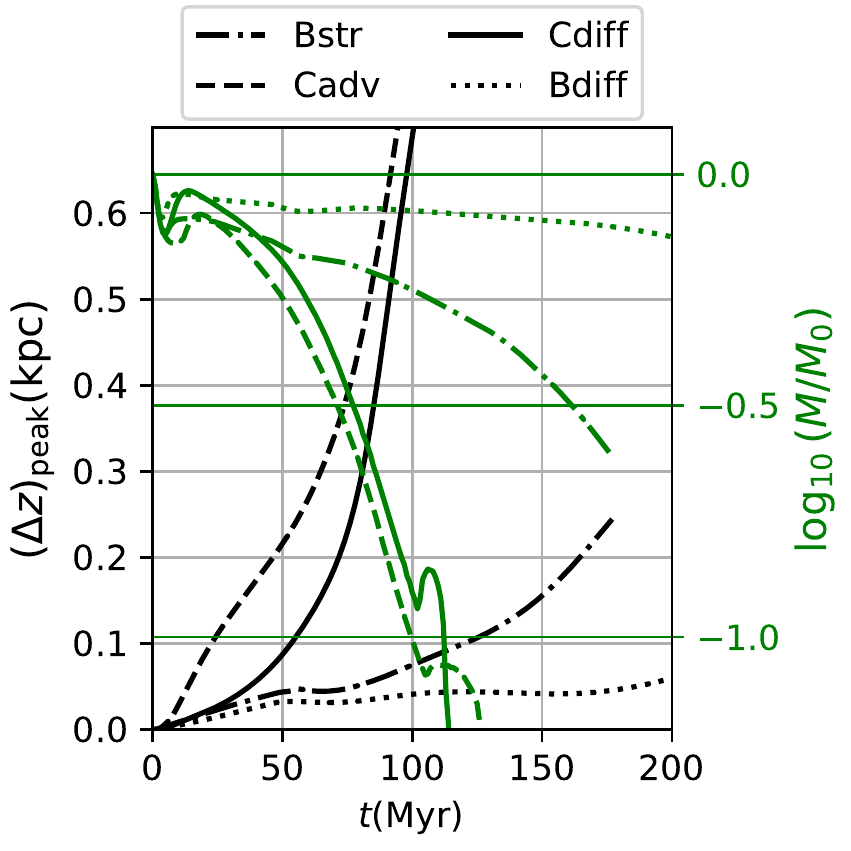}
    \caption{Plot of maximum tube height (black) and average mass (green) within \review{$x = \pm 250 \,\mathrm{pc}$ of the injection location} for simulations related to Q2.  Note the vertical axis has been extended compared to Figure \ref{fig:AllSims}. \review{Solid lines show simulation \texttt{Cdiff}, dashed lines show simulation \texttt{Cadv}, dash-dotted lines show simulation \texttt{Bstr}, and dotted lines show simulation \texttt{Bdiff}.} The streaming case has the flux tube rise buoyantly, but only after a long time $\gtrsim 100 \,\mathrm{Myr}$. In contrast, the advection and diffusion dominated cases rise early $\lesssim 100 \,\mathrm{Myr}$ and cause a significant amount of mass loss from the flux tube. \review{ The streaming case is primarily inhibited by the magnetic field strength, not the streaming transport, because in simulation \texttt{Bdiff} the flux tube takes an even longer time to rise.} }
    \label{fig:Q2}
\end{figure}

We ran four simulations to determine the impact of cosmic ray transport: simulation \texttt{Bstr} is streaming dominated, simulation \texttt{Cadv} is advection dominated, \texttt{Cdiff} is diffusion dominated, \review{and \texttt{Bdiff} is diffusion dominated with a strong magnetic field.} With these simulations, it is clear that streaming does a poor job of launching material when compared to diffusion and advection, which both disrupt the ISM on a short time scale $\sim 100 \,\mathrm{Myr}$. \boldreview{However, \texttt{Bdiff} shows this delay is not a result of the cosmic ray transport by streaming. Instead, the large magnetic field (necessary for streaming to be dominant) resists any bending created by the injection. Streaming is more effective than diffusion at disrupting the vertical structure when magnetic tension is a dominant force.} Figure \ref{fig:Q2} shows the flux tube averaged mass and height for these three simulations.

Simulation \texttt{Bstr} does not start to rise buoyantly until a significant amount of mass has been lost from the flux tube. In comparison, the buoyant rise in simulation \texttt{Cdiff} happens quickly \review{(see panels of Figure \ref{fig:AllSims})}. The advective simulation \texttt{Cadv} is mainly driven by an explosive launching, instead of buoyancy. However, it and the diffusion case produce similar results in terms of flux tube movement. The difference between those two simulations is more apparent in the beginning, when the launching is different. 

Of these simulations, the diffusion case is the most surprising. \review{Initially, one might think cosmic ray diffusion may have a smaller effect on feedback processes than streaming and advection, because there is less time for cosmic rays to impact the ISM (See Equation \ref{eqn:timeDiff}) when using the Milky Way value of the diffusion coefficient.} However, our simulation \texttt{Cdiff} shows a large cosmic ray injection can generate enough force through the cosmic ray pressure gradient to move mass out of the disk and bend the magnetic field. 

\subsection{Dependence on Injection Characteristics (Q3 \& Q4)}
\label{sec:Results:Q34}

\begin{figure}
    \centering
    \includegraphics[width=0.48\textwidth]{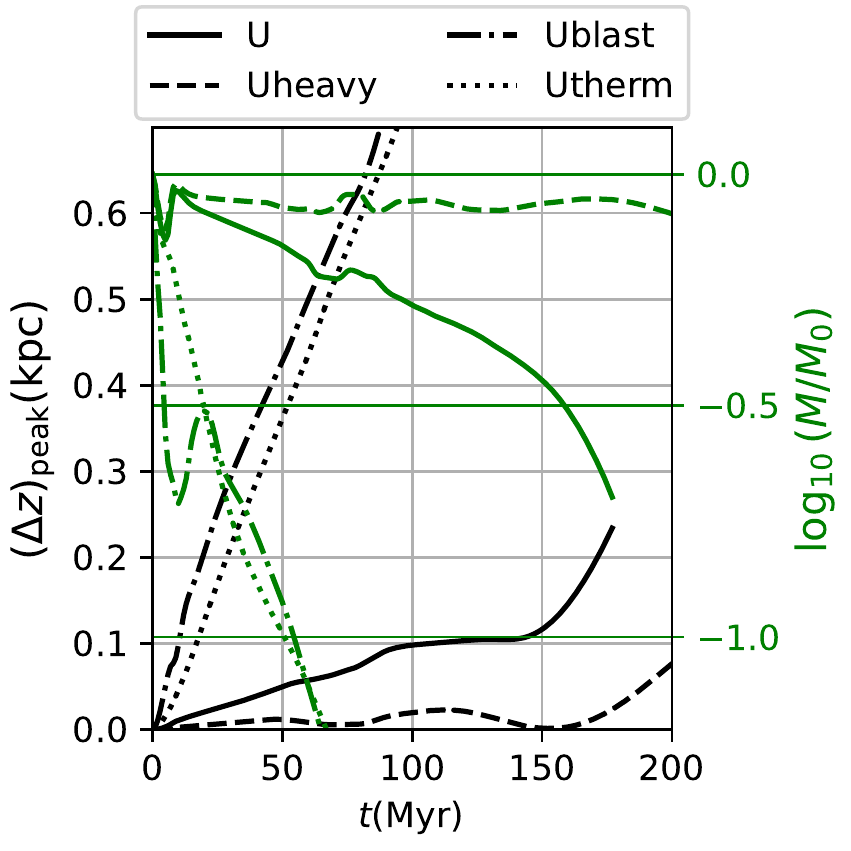}
    \caption{Plot of maximum tube height (black) and average mass (green) within $x = \pm 250 \,\mathrm{pc}$ of the injection for simulations related to Q3 and Q4. Note the vertical axis has been extended compared to Figure \ref{fig:AllSims}. \review{Solid lines show simulation \texttt{U}, dashed lines show simulation \texttt{Uheavy}, dash-dotted lines show simulation \texttt{Ublast}, and dotted lines show simulation \texttt{Utherm}.} Simulation \texttt{Uheavy} does eventually rise from the disk after $200\,\mathrm{Myr}$. Simulation \texttt{Ublast} drives an explosive launching faster than the thermal injection in \texttt{Utherm}.}
    \label{fig:Q34}
\end{figure}

We ran three simulations focused on the injection characteristics. The first, \texttt{Uheavy}, placed the injection lower in the galactic disk. While the flux tube rose later than in other simulations, this simulation gave similar results when compared to \texttt{U}. Simulation \texttt{Ublast} examined how increasing the injection energy to $10^{52} \,\mathrm{erg}$ would change the dynamics. This large injection caused significant change in the ISM, causing the flux tube to rise rapidly. \review{This simulation suggests that future simulations, with multiple injections, could lead to rapid disruption.} In Simulation \texttt{Utherm}, we replaced the cosmic ray injection with a thermal injection of similar magnitude. This injection contained no mass, but was an over-pressure $(T\sim10^5\,\mathrm{K} > T_0 \sim 10^{4}\,\mathrm{K})$ region. This thermal energy injection caused rapid mass flow off the flux tube, driving a buoyant rise. \review{However, the simulation is physically unrealistic because we do not include radiative cooling (see Section \ref{sec:Results:Utherm}).} The tube averaged mass and height of these simulations are shown as part of Figure \ref{fig:Q34}.

\noindent
\begin{figure*}
    \centering
     \includegraphics[width=0.48\textwidth]{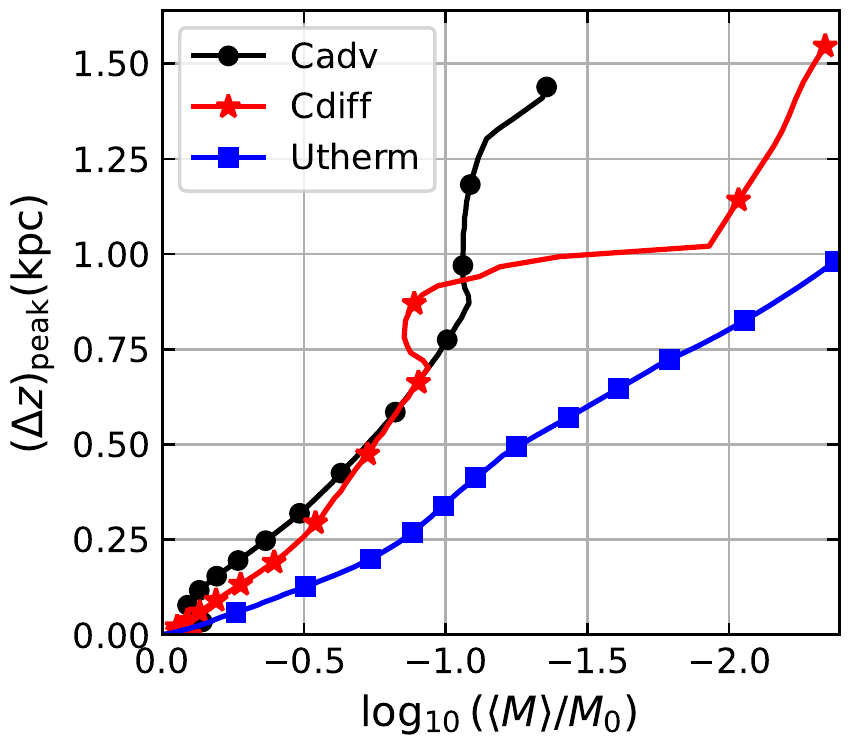}\hfill
    \includegraphics[width=0.48\textwidth]{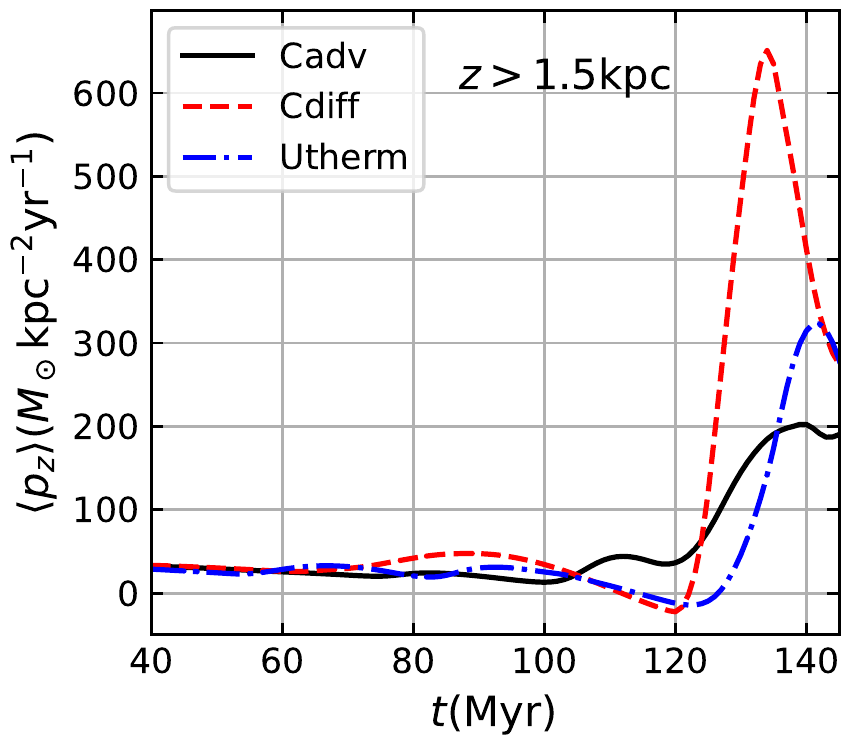}
    \vspace{0.1in}
    \includegraphics[width=0.98\textwidth]{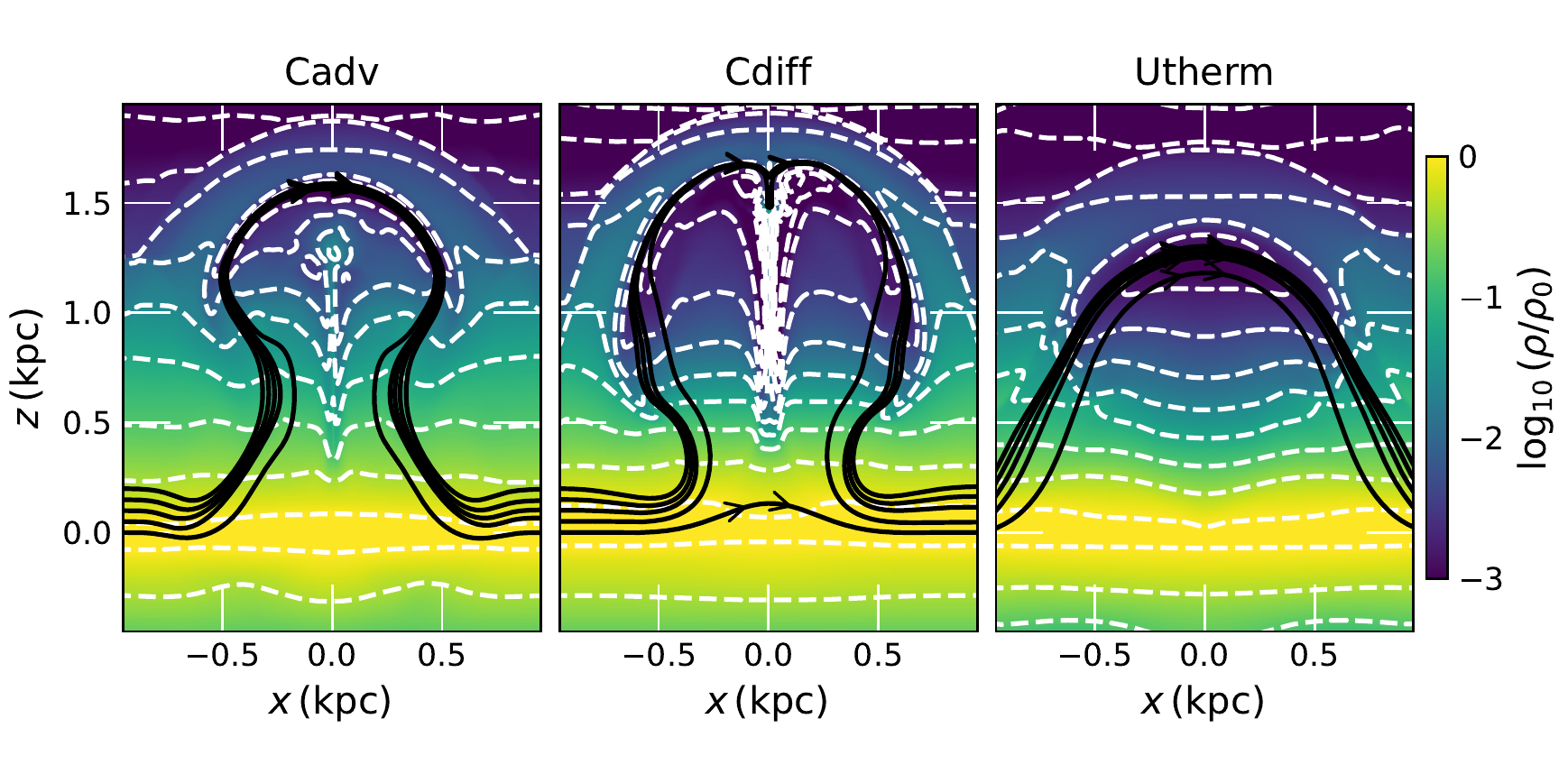}
    \caption{ \textit{Top Row, left plot: }The evolution over time of each flux tube's height and mass. \review{Solid lines and filled markers correspond to the tube's peak height.} Each simulation is shown until $130\,\mathrm{Myr}$, with markers placed at $10\,\mathrm{Myr}$ increments. Black circles show simulation \texttt{Cadv}, red stars show simulation \texttt{Cdiff}, and blue squares show  simulation \texttt{Utherm}. While the thermal injection in simulation \texttt{Utherm} initially moves material up faster, at late times the cosmic ray injections in simulations \texttt{Cadv} and \texttt{Cdiff} overtake simulation \texttt{Utherm}. After $130\,\mathrm{Myr}$, the diffusion transported cosmic ray injection has created the \review{most vertically extended flux tube.} \review{ \textit{Top Row, Right Plot:} For the same simulations, we show the average momentum in the vertical direction in the region of $z>1.5 \,\mathrm{kpc}$. \texttt{Cdiff} creates the largest mass flux. The decrease around $140\,\mathrm{Myr}$ is when the flux tube leaves the simulation through the upper boundary.} \textit{Bottom Row: }Cuts of simulations \texttt{Cadv}, \texttt{Cdiff}, and \texttt{Utherm}. These simulations were the fastest at \review{disrupting the vertical structure, with each image shown at a time of $130\,\mathrm{Myr}$}. The black lines are magnetic field lines and the white dashed lines are density contours. These plots are cuts of our 3D simulations at $y=0$. \review{Simulation \texttt{Cdiff} reaches the highest point, with significantly adjusted magnetic field structure.} }
    \label{fig:Fast}
\end{figure*}

\subsection{Fast Disruption: Cosmic rays vs. Thermal Injection}
\label{sec:Results:Fast}
The simulations which produced the largest changes in the shortest time for an injection energy $10^{51}\,\mathrm{erg}$ were \texttt{Cadv}, \texttt{Cdiff}, and \texttt{Utherm}. \review{These injections had flux tubes rise faster than a traditional Parker Instability, and caused the flux tube to lose approximately half its mass in $\lesssim 100 \, \mathrm{Myr}$. In the top left panel of Figure \ref{fig:Fast} we show the evolution of peak tube height against mass. The markers on each line denote $10\,\mathrm{Myr}$ steps, whereas each line is made with points at a resolution of $1\,\mathrm{Myr}$.} The final marker is at $130\,\mathrm{Myr}$ for each simulation. Even though the cosmic ray injection simulations take a longer time to start rising, they rise at a faster rate than the thermal injection. They overtake the thermal injection after approximately $90\,\mathrm{Myr}$, and they have more mass at that time. The density at $y=0$ of the three simulations at this final time, $130\,\mathrm{Myr}$, is shown in the bottom panel of Figure \ref{fig:Fast}. Each simulation produces a different magnetic field structure. Considering simulation \texttt{Utherm} is an overestimate of thermal injection's impact (see cooling time argument in Section \ref{sec:Results:Utherm}), we can focus on the difference between \texttt{Cadv} and \texttt{Cdiff}. The left and middle plots in the lower row of Figure \ref{fig:Fast} show that cosmic ray transport changes the both the flow of gas around the rising flux tube and the shape of the magnetic field lines. \review{The top right plot of Figure \ref{fig:Fast} shows the average vertical momentum of the cells with $z > 1.5\,\mathrm{kpc}$ for each of the three simulations. Following the red dashed line for simulation \texttt{Cdiff}, it is clear the diffusion dominated simulation creates the largest mass flux from the disk. While this gas may fall down given enough time and vertical expansion distance, we are unable to continue to follow that gas because it leaves out of the top of our simulation $(z = 2\,\mathrm{kpc})$. }

\section{Conclusion}
\label{sec:Conc}
We ran ten simulations of cosmic ray injection into a vertically stratified medium, using the \texttt{Athena++} code. These simulations illustrated the effect of cosmic ray injection in a galactic disk, on intermediate scales $(\sim 1 \,\mathrm{kpc})$ larger than the ISM's fine structure and smaller than the entire galaxy. By exploring an extended parameter space, we produced the highest resolution picture yet of localized cosmic ray injection on these mesoscales. We also showed that cosmic ray transport dictates the impact of cosmic rays on the ISM. Below, we provide the key points and results of this work:
\begin{itemize}
    \item Cosmic ray diffusion of a cosmic ray injection can change the ISM's vertical structure and a galaxy's magnetic field on timescales $<100 \, \mathrm{Myr}$.
    \item \boldreview{The large magnetic field strength necessary for cosmic ray streaming dominance over diffusion limits the rapid disruption of the flux tube because of increased magnetic tension. Streaming is effective at disrupting the vertical structure, but takes a longer time than in weak magnetic field cases.}
    \item A flux tube disrupted by cosmic ray injection will rise faster at later times than one disrupted by thermal injection, producing a larger mass flow of material out of the galactic disk. 
\end{itemize}

Our simulations provided useful heuristic results by considering cosmic ray injection in a stratified medium. These simulations are clearly limited in their realism because galactic disks are not uniformly stratified. The ISM in galactic disks is multiphase and clumpy. Additionally, by neglecting the dynamics of the stellar gravitational potential, we lack forcing terms which could change the effects of these energy injections. Future work may need to focus on diffusion as the primary cosmic ray transport mechanism, along with implementation of more realistic ISM conditions (multiple gas phases,  galactic rotation, etc.). The non-constant $\alpha$, $\beta$ of the injection have a significant impact. This non-uniformity should be extended to the background medium, instead of using the constant $\alpha,\beta$ assumption originally introduced by \cite{1966Parker}. \review{Variable magnetic field strength is particularly important, as it could amplify the importance of cosmic ray streaming in low $\beta_\mathrm{pl}$ regions.}

In these simulations, we neglected radiative heating and cooling, but those processes are an important consideration in the ISM. The radiative cooling would be particularly important for the gas heated by cosmic ray streaming. Cooling will minimize, and possibly remove, the impact of the thermal injection in simulation \texttt{Utherm}. \boldreview{Additionally, having only a single cosmic ray injection by supernovae in our simulation volume over a time of $100\,\mathrm{Myr}$ is unrealistic.} In future work, we aim to include multiple injections at different times and locations. Separating the injections in space opens up a variety of other situations which make it difficult to isolate the buoyancy the injection creates, hence why we only considered a single injection location in this work. 

For large scale (cosmological or galactic) simulations, there needs to be some consideration of cosmic ray injection. Even simulations which include cosmic rays generally do not resolve their injection into the ISM. Our work shows a cosmic ray injection can drive an upward flow in $<100\,\mathrm{Myr}$ after their injection, and most of that upward movement actually happens in $<20\,\mathrm{Myr}$. \review{A resolution of $\sim 100 \,\mathrm{pc}$ would be enough to illustrate the impact of an upward flow along the magnetic field (consider bottom row of Figure \ref{fig:Fast}). A sub-grid physics module related to cosmic ray injection could initiate this flow before letting it evolve independently.} In the $y$ direction, across the magnetic field and in the plane of the disk, higher resolution would be necessary. The width of the flux tube barely reaches $100 \,\mathrm{pc}$ in that direction, and the overall dynamics in that direction are minimal. The main impact the $y$ direction had in our simulations is to allow mass above the flux tube to move out of the path of the rising flux tube. The actual resolution of that dimension is less significant (See Appendix \ref{app:Bound}). 

Our work also shows cosmic ray injection is an important part of dynamics in a galaxy. Moving, heating, and compressing gas all have an impact on where stars form and on galactic structure. Since the eventual rise of the flux tube happens in a short time  $(<20\,\mathrm{Myr})$, the dynamics created by cosmic ray injection are less likely to be washed out by galactic rotation and other dynamical processes. Our results also shed light on the respective roles of cosmic ray compressibility and buoyancy alluded to in Section \ref{sec:Back:Parker}. \review{Finally, our results suggest that including effects of localized cosmic ray injection in global simulations could be a necessary step in accurately modeling galactic outflows and evolution.}

\section{Acknowledgements}
We would like to thank Yan-Fei Jiang for sharing the code for cosmic rays in \texttt{Athena++}, originally presented in \cite{2018Jiang}. \review{We are very thankful for the comments provided by an anonymous reviewer. Their suggestions significantly improved this work.} We would like to thank Evan Heintz and Chad Bustard for helpful discussions about our simulations and the Parker instability. \review{We also appreciate commentary from Ryan Farber and Cassi Lochhaas which improved the paper.}  This work was funded by NSF grant AST-2007323 and NASA FINESST Grant 80NSSC22K1749.


\software{Athena++ \citep{2020Stone,2018Jiang}, \review{ MatPlotLib \citep{2007Matplotlib}, NumPy \citep{2011NumPy,2020NumPy}, AstroPy \citep{2013AstroPy,2018Astropy}}}

\bibliography{sample631}{}
\bibliographystyle{aasjournal}

\appendix 
\begin{figure}[b]
    \centering
    \includegraphics[width=0.98\textwidth]{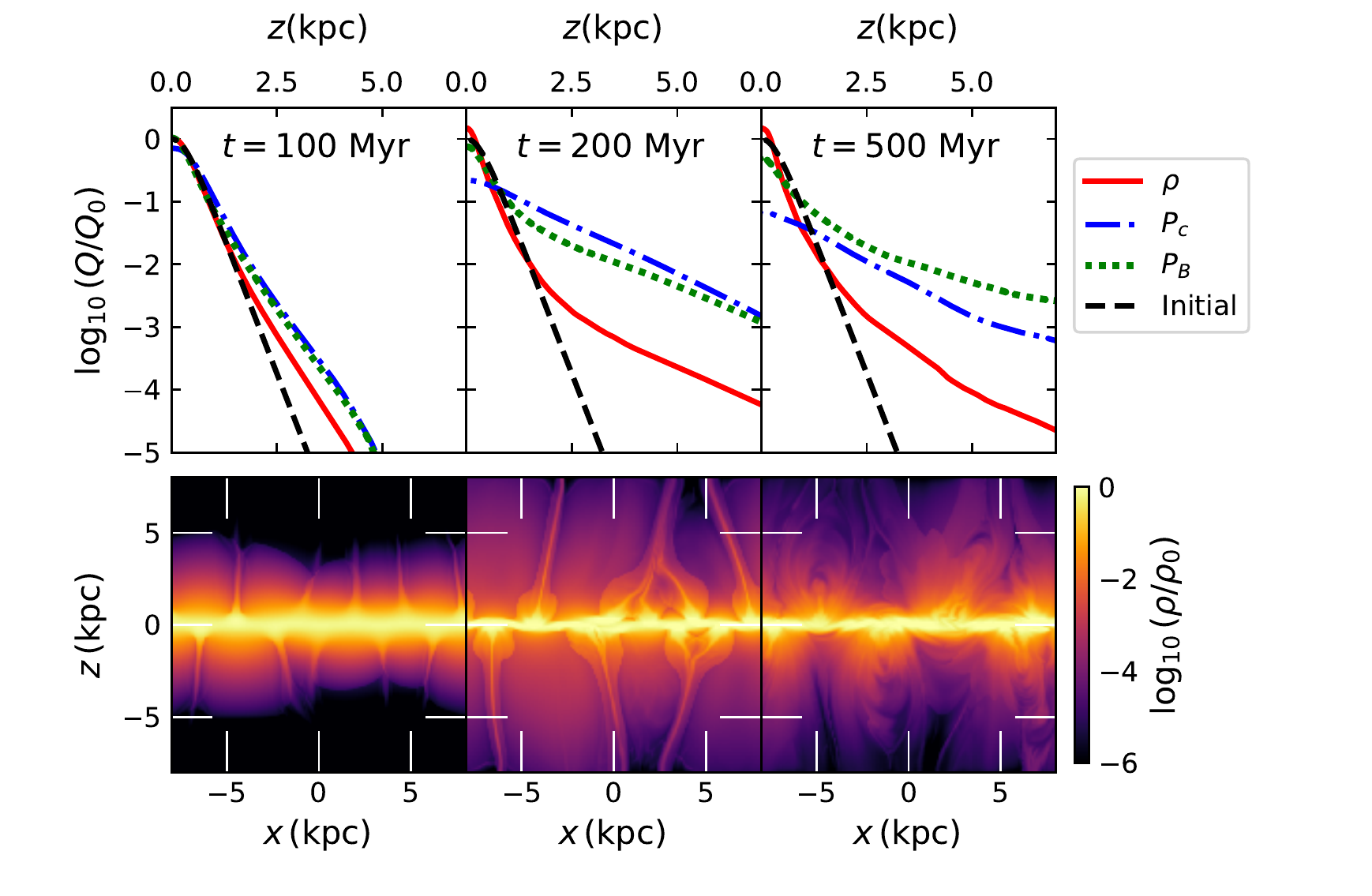}
    \caption{\review{\textit{Top row:} Vertical profiles of horizontally averaged gas density (solid red line), cosmic ray pressure (dash-dotted blue line), and magnetic pressure (dotted green line). The initial vertical profile for the quantities is shown as well (dashed black line). The initial profile is the same for each quantity because we normalize by the initial midplane value of each quantity, and each physical variable is proportional to the other (see Section \ref{sec:Back}). \textit{Bottom row:} Gas density at corresponding time dump to the vertical profile above each image. The total mass in the simulation is conserved, with a loss of $\sim 0.1\%$ through the diode boundary conditions in the vertical $(\hat{z})$ directions.}}
    \label{fig:Parker2D}
\end{figure}

\section{Comparison with Parker Instability Simulations}
\label{app:Compare}
To test our numerical methods and initial conditions, we ran a 2D simulation matching one in \cite{2020Heintz}. Those simulations used the FLASH code \citep{2000Fryxell}, along with the streaming transport method in \cite{2010Sharma}. Matching those parameters (originally based on \cite{2016Rodrigues}), we use 
$\rho_0 = 6.76\cdot 10^{-25} \, \mathrm{g}\, \mathrm{cm}^{-3}$, 
$P_{g,0} = 8.19 \cdot 10^{-13}\,\mathrm{erg}\,\mathrm{cm}^{-3}$,
$g_*=2 \cdot 10^{-9} \, \mathrm{cm}\, \mathrm{s}^{-2}$,
$\eta=2$, 
$H=250\,\mathrm{pc}$,
$\alpha=10/3$, $\beta = 1.25$, and $V_m=0.01c$.

We find that that our simulation method is consistent with that of \cite{2020Heintz}, and our results are illustrated in Figure \ref{fig:Parker2D}. \review{ This figure shows horizontally $(\hat{x})$ averaged profiles of the gas density, cosmic ray pressure, and magnetic pressure at several time dumps, compared with the original profile for the quantities. Below the averaged profiles, we show the gas density at each selected time dump.} The magnetic and cosmic ray pressures decrease more slowly away from the plane as the magnetic field becomes bent, while gas is compressed towards the midplane. Gravity pulls the gas down along the magnetic flux tubes, which have turned perpendicular to the disk in several locations. Our simulation matches expected behavior (see Figure 11 of \cite{2020Heintz}, a similar plot) and evolves on a similar time scale to \cite{2022Tharakkal}. \boldreview{These profiles illustrate the transition from a linear growth regime to a nonlinear regime examined in depth by \cite{2022Tharakkal}. Additionally, these 2D simulations illustrate the cosmic ray method implemented in \texttt{Athena++} by \cite{2018Jiang} is useful in studying the Parker instability, since it is in reasonable agreement with other numerical simulations.}

\section{Boundary Conditions \review{and Dimensionality}}
\label{app:Bound}

In the direction of stratification $(\hat{z})$, we use vacuum (also known as diode) boundary conditions. These conditions make it impossible for inflow to occur, since the boundary cells are set to the density and pressure floor of the numerical simulation. We also avoid setting up a steep discontinuity at this boundary by extending the simulation several scale heights in the vertical direction. Because of this extension, the cells near the boundary are already almost at the density and pressure floors when the simulation starts. Then, any dynamical activity beyond that initial equilibrium profile will propagate out of the simulation. 

\begin{figure}
    \centering
    \includegraphics[width=0.98\textwidth]{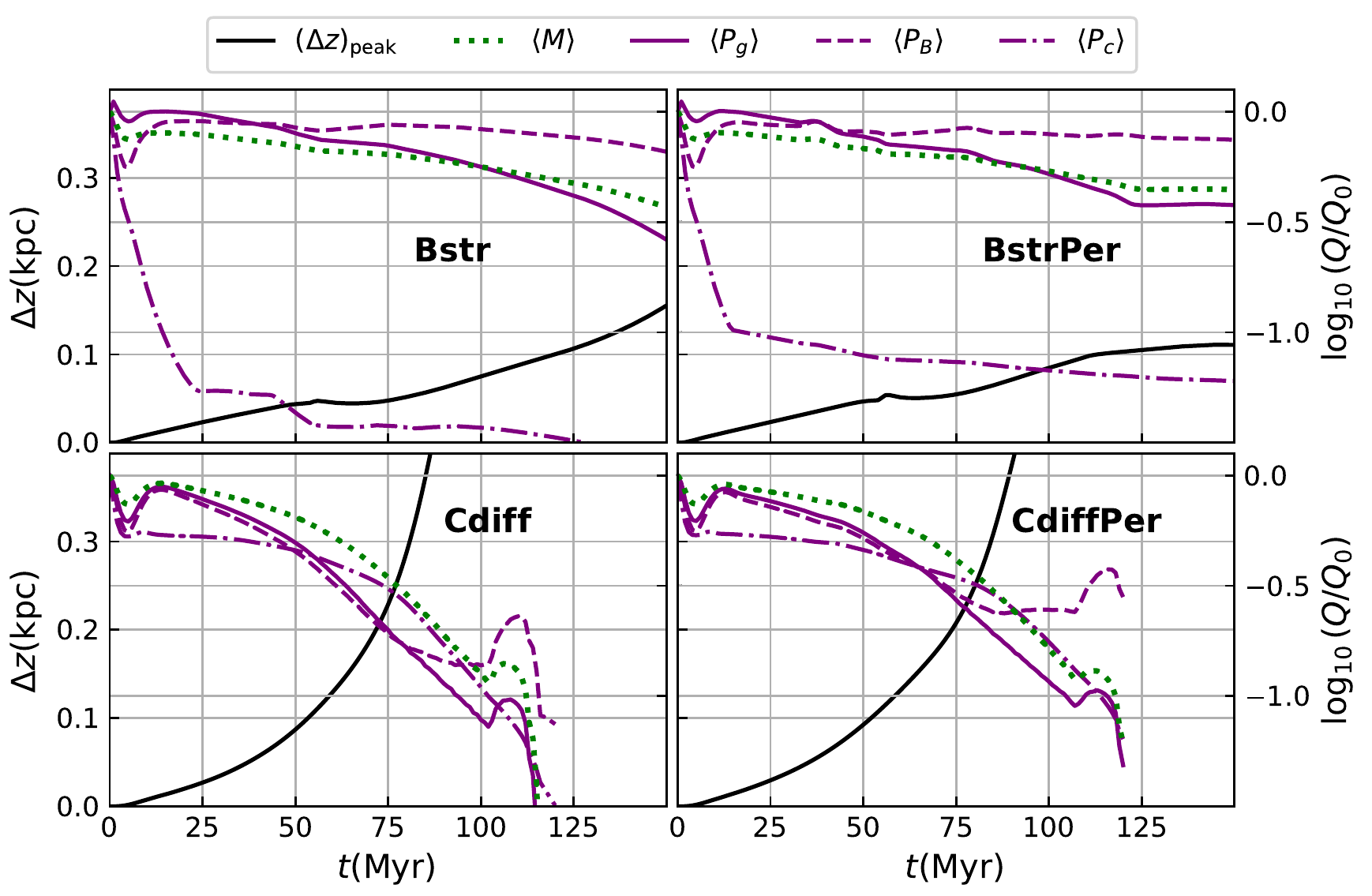}
    \caption{\review{Plots matching Figure \ref{fig:AllSims}, but comparing simulations \texttt{Bstr} and \texttt{Cdiff} to identical simulations with periodic boundary conditions, \texttt{BstrPer} and \texttt{CdiffPer}. For each simulation, the solid black line is maximum height of the flux tube at a given time. The green dotted line is the average mass in the region $x<|250\,\mathrm{pc}|$, as a fraction of the initial average mass. The purple solid, dashed, and dash-dotted lines are the gas, magnetic, and cosmic ray pressure, respectively, averaged in the same region. The second $y$ axis on the right side applies to the lines of mass and pressures, with each quantity $Q \in \{ \langle M \rangle, \langle P_g \rangle , \langle P_B \rangle, \langle P_c\rangle\}$ normalized by either $P_0=10^{-12}\,\mathrm{erg}\,\mathrm{cm}^{-3}$ or the initial value (for mass $\langle M \rangle$).}}
    \label{fig:Per}
\end{figure}

In directions in the plane of the disk ($\hat{x}$ and $\hat{y}$) we use outflow boundary conditions. While these allow inflow from boundary cells after the crest of a wave passes the boundary, there is very little error if the gas is quickly moving out of the simulation at that boundary. Our current problem satisfies this assumption because the flow is either static, or large (when the perturbation pushes gas down a magnetic flux tube and toward the boundary). Ideally, we would again use vacuum boundary conditions to stop inflow. However, we cannot use a vacuum in this direction since this would create a significant discontinuity near the midplane of the simulation, where there are gas densities and pressures above the floor values. \review{Therefore, outflow boundary conditions are the best ones for treating a single, spatially and temporally isolated, injection.}

With the goal of minimizing numerical errors, simulations often exploit periodic boundary conditions. \boldreview{In the current work, we wanted to avoid tying down the flux tubes at the boundaries of the simulation in this study, because that can have a stabilizing effect \citep{1992Zweibel}. However, to make sure our simulation results were not significantly different when using periodic boundaries, we ran two simulations, \texttt{BstrPer} and \texttt{CdiffPer}, with periodic boundary conditions but the same initial conditions as \texttt{Bstr} and \texttt{Cdiff}, respectively. The flux tube dynamics of those simulations are shown in Figure \ref{fig:Per}. Simulations \texttt{Cdiff} and \texttt{CdiffPer} show very similar results. For simulations \texttt{Bstr} and \texttt{BstrPer}, the difference is larger: the periodic boundary conditions tie down the magnetic flux tube at the boundaries, making it more difficult for the tube to rise after $100\,\mathrm{Myr}$. Overall, the trends are the same, but the timing is delayed when using periodic boundaries. }

\begin{figure}
    \centering
    \includegraphics[width=0.98\textwidth]{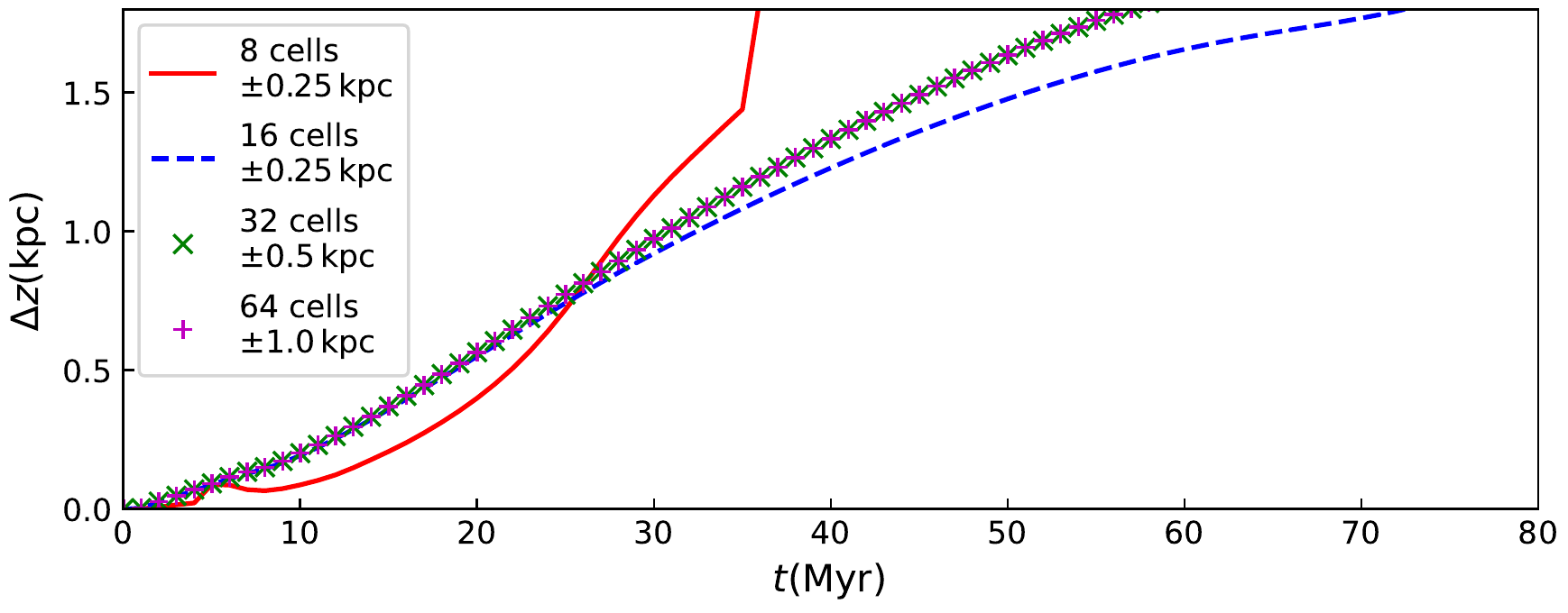}
    \caption{\review{Height of flux tube in simulations with large injection energy $E_\mathrm{inj} \approx 10^{51}\,\mathrm{erg}$, similar to simulation \texttt{Ublast}. These simulations were run to examine the impact of the third dimension, $\hat{y}$, on our results. The red solid line shows a simulation with only $8$ numerical cells in the $\hat{y}$ direction, the blue dashed line used $16$ numerical cells, the green X markers are from a simulation with $32$ numerical cells, and the purple plus markers are from a simulation with $64$ numerical cells. The simulations with more room in the $\hat{y}$ direction are well converged at a length of $\pm 0.5\,\mathrm{kpc}$. We chose to use $50$ numerical cells in our production runs, with a resolution of $\Delta y = 20 \,\mathrm{pc}$. } }
    \label{fig:Dims}
\end{figure}

\review{We also examined the effect of dimensionality on our simulations. Initial 2D simulations differed from our results in 3D simulations significantly, with tubes taking much longer to rise in 2D simulations. This delay is caused by flux tubes above our injection being unable to move out of the way of the rising flux tube where the injection took place. This effect is also apparent in Figure \ref{fig:Dims}, where we examine different resolutions and numerical sizes in the third dimension, $\hat{y}$. For $8$ cells in the $\hat{y}$ direction and resolution $\Delta y = 62.5\,\mathrm{pc}$, the tube slowly rises before bursting upward. This trend is similar to our 2D simulations, whose initial rise takes longer. In these simulations, the rise only happens quickly because we used a large injection energy $E_\mathrm{inj} \approx 10^{51}\,\mathrm{erg}$ similar to simulation \texttt{Ublast}. For $16$ cells and a resolution $\Delta y = 31.25\,\mathrm{pc}$, we see a smooth well-behaved flux tube rise. By increasing the width of the box to $32$ and $64$ numerical cells with the same resolution as the $16$ cell run, we get clear convergence of results. From these simulations, we chose a width of $\pm 0.5 \,\mathrm{kpc}$ and resolution of $\Delta y = 20\,\mathrm{pc}$, resulting in $50$ numerical cells.}

\begin{figure}
    \centering
    \includegraphics[width=0.98\textwidth]{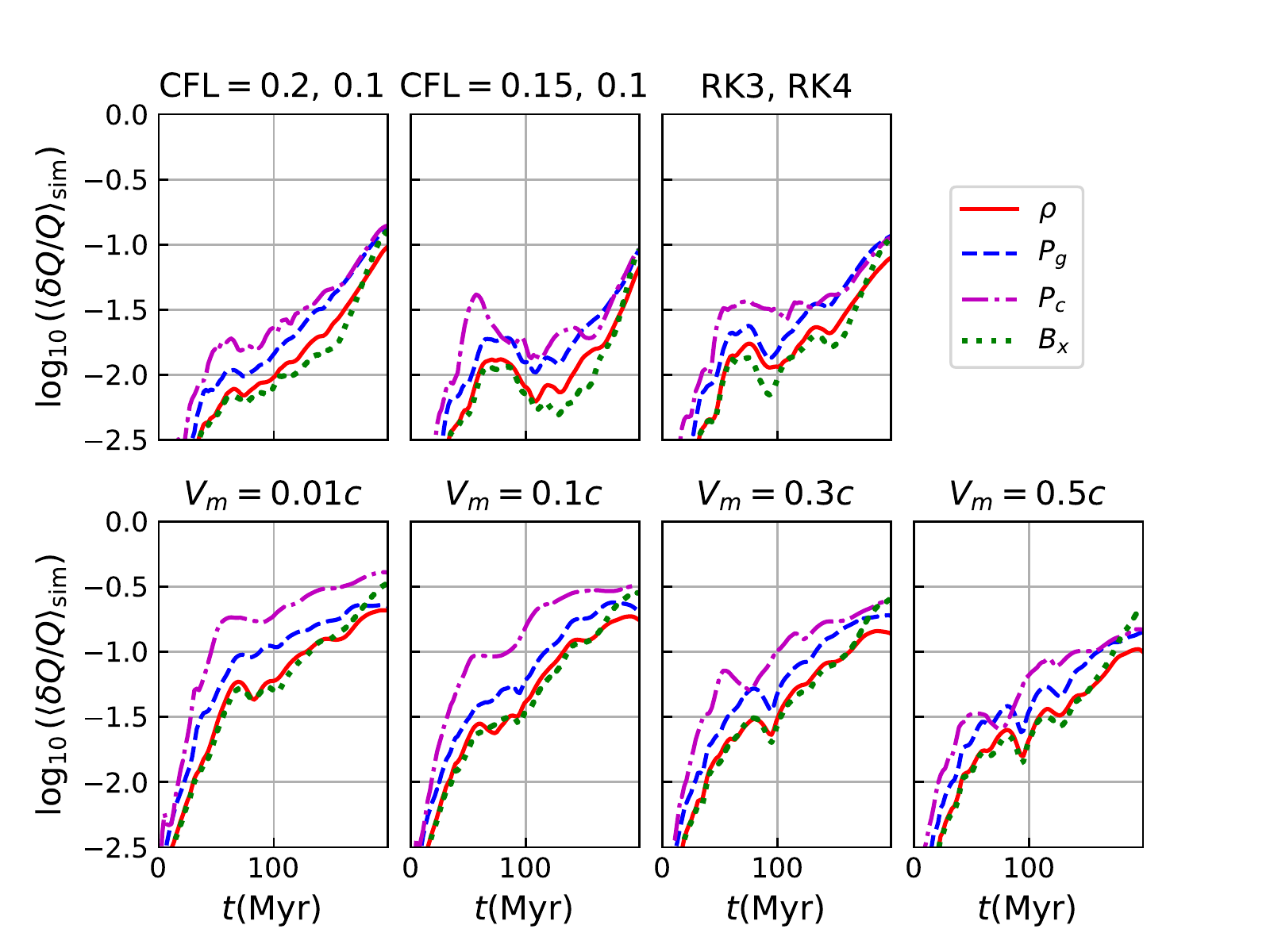}
    \includegraphics[width=0.98\textwidth]{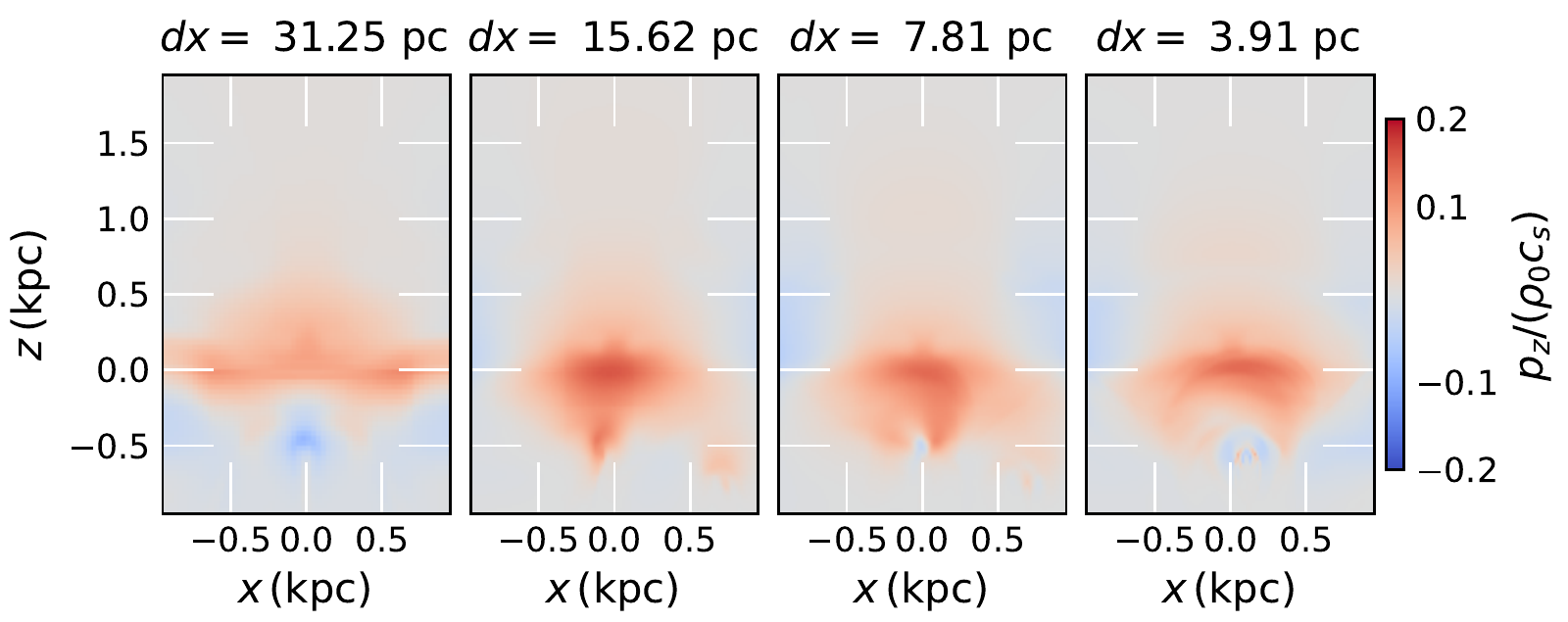}
    \caption{\review{Convergence of key numerical parameters. \textit{Top row, first plot:} Average cell-by-cell numerical error between $\mathrm{CFL} = 0.2$ and $\mathrm{CFL} = 0.1$. \textit{Top row, second plot:} Average cell-by-cell numerical error between $\mathrm{CFL} = 0.15$ and $\mathrm{CFL} = 0.1$. \textit{Top row, third plot:} Average cell-by-cell numerical error between a run with a third order Runge-Kutta time integrator and a run with a fourth order Runge-Kutta time integrator. \textit{Middle row:} Average cell-by-cell numerical error between runs with varying $V_m$. Each has error measured with respect to a $V_m=c$ simulation. \textit{Bottom row:} Plot of vertical momentum $p_z$ in simulations with different resolutions $dx = dz$.}}
    \label{fig:ConvRKTINT}
\end{figure}

\section{Simulation Convergence}
\label{app:Convergence}
To mitigate the significance of purely numerical parameters we performed several convergence tests. \review{We used 2D simulations to narrow our choice of time integrator, Courant-Friedrich-Lewy (CFL) number, resolution, and the reduced speed of light constant $V_m$ (see Equation \ref{eqn:JiangCRMomentum}).} Using 2D simulations allowed us to save computational resources while exploring the convergence of these simulations. We illustrated in Section \ref{app:Bound} that 2D simulations would be less accurate because they limit the movement of magnetic flux tubes, but most of the fast, dynamic flow is still in the $xz$-plane. For numerical parameters, the motion in that plane is where we need be concerned. All the convergence tests here used a weak injection of $10^{50}\,\mathrm{erg}$ with other parameters equal to those of simulation \texttt{U}. With the weaker injection and 2D restricted motion, the flux tube  begins to rise buoyantly by $200\,\mathrm{Myr}$.

The time integrator choice of 3rd order Runge-Kutta and CFL number ($0.2$) were well converged, \review{as the average cell-by-cell error is under $10\%$ in the top row of Figure \ref{fig:ConvRKTINT}. The error associated with the modified speed of light is larger: in the middle row, the cell-by-cell error for various values of $V_m$ are shown, each compared to a simulation with $V_m = c$. From these, we pick $V_m = 0.3c$ because there is not a huge increase in accuracy by going to higher values of $V_m$. Increasing $V_m$ beyond $0.3c$ would decrease the timestep below $0.1 \,\mathrm{kyr}$, increasing computational resource requirements without an equivalent increase in accuracy. Finally, in the bottom row of Figure \ref{fig:ConvRKTINT}, we show the vertical momentum $p_z$ for simulations with varying resolution. Instead of comparing these simulations via interpolation of the high resolution simulations, we focus on how similar the dynamics are between each simulation. Each simulation had an aspect ration of $1$, with resolution $\Delta x = \Delta z$. Any higher resolution than $15 \,\mathrm{pc}$ appears to reproduce a similar vertical momentum on the flux tube. Also, the structures are similar between resolutions $15.62\,\mathrm{pc}$, $7.81\,\mathrm{pc}$, and $3.91\,\mathrm{pc}$. From these, we decided to run our perfomance simulations with a resolution $\Delta x = \Delta z = 10 \,\mathrm{pc}$. }

\end{document}